\documentclass[onecollarge]{svjour2}   
\pdfoutput=1 

\smartqed  

\usepackage{graphicx}
\usepackage{amsmath}
\usepackage{amssymb}
\usepackage{color}
\usepackage{cite}

\newcommand{\sref}[1]{Section \ref{#1}}
\newcommand{\fref}[1]{Figure \ref{#1}}
\newcommand{\aref}[1]{Appendix \ref{#1}}

\newcommand{\erf}[1]{\mathrm{erf}\!\left( {#1} \right)}
\newcommand{\erfc}[1]{\mathrm{erfc}\! \left( {#1} \right)}
\newcommand{\X}{\mathbf{X}}
\newcommand{\Y}{\mathbf{Y}}
\newcommand{\arctanh}{\mathrm{arctanh}}

\journalname{J. Stat. Phys.}

\begin{document}

\title{Tagged particle in single-file diffusion}

\author{P L Krapivsky \and Kirone Mallick         \and
        Tridib Sadhu
}

\institute{P L Krapivsky \at
              Physics Department, Boston University, Boston,
              Massachusetts 02215, USA
           \and
           Kirone Mallick and Tridib Sadhu  \at
          Institut de Physique Th\'{e}orique, CEA/Saclay, F-91191 Gif-sur-Yvette Cedex, France
}

\maketitle

\begin{abstract}
Single-file diffusion is a one-dimensional interacting infinite-particle system in which the order of particles never changes. An intriguing feature of single-file diffusion is that the mean-square displacement of a tagged particle exhibits an anomalously slow sub-diffusive growth. We study the full statistics of the displacement using a macroscopic fluctuation theory. For the simplest single-file system of impenetrable Brownian particles we compute the large deviation function and provide an independent verification using an exact solution based on the microscopic dynamics. For an arbitrary single-file system, we apply perturbation techniques and derive an explicit formula for the variance in terms of the transport coefficients. The same method also allows us to compute the fourth cumulant of the tagged particle displacement for the symmetric exclusion process. 

\end{abstract}

\section{Introduction \label{intro}}

In non-equilibrium statistical mechanics, dynamical properties of
interacting many-body systems play as important role as
non-equilibrium steady states. Unlike the equilibrium properties,
dynamical properties depend on the initial and boundary conditions and
on details of the interactions.

The motion of an individual particle in a system of interacting particles is a fundamental {\it dynamical} problem in statistical mechanics even when the entire system is in equilibrium, or in a non-equilibrium steady state. A single Brownian particle exhibits normal diffusion---the mean-square displacement grows linearly with time. In a gas composed of impenetrable diffusing particles, the mobility of an individual particle is reduced due to the presence of other particles. In two and three dimensions, this effect can be significant in dense gases, but qualitatively the tagged particle still undergoes normal diffusion. In one dimension, however, the order of particles is preserved and transport can become anomalously slow. Transport in such systems is known as single-file diffusion.

Single-file diffusion is prevalent in numerous physical, chemical and biological processes:  molecular motion inside porous medium like zeolite \cite{CHOU1999,KARGER1992}, water transport inside a carbon nanotube \cite{Das2010}, motion of tagged monomers in a polymer chain \cite{Gupta2013}, sliding of proteins in a DNA sequence \cite{Li2009}, ion channels through biological membranes \cite{HODGKIN1955}, super-ionic conductors \cite{Richards1977}, etc.

The motion of an individual (tagged) particle strongly depends on whether the dynamics is biased or not. A diffusive behavior emerges when the dynamics is biased \cite{Demasi1985,Kipnis1986,Majumdar1991,Beijeren1991,Ferrari1996,Liggett2004,ShamikSatya}. In the unbiased case, the variance of the displacement $X_T$ exhibits a sub-diffusive $\sqrt{T}$ growth. In this paper, we focus on the unbiased situation.\footnote{If the particle dynamics is ballistic, the tagged particle exhibits a normal diffusion behavior \cite{Percus1974,Abhishek2013}.}

The sub-diffusive $\left \langle X_T^2 \right\rangle\sim \sqrt{T}$ behavior was first derived by Harris  \cite{Harris1965} for the system of Brownian particles interacting only through the hard-core repulsion; the same qualitative behavior was found to occur \cite{Arratia1983,Sethuraman2013} for the symmetric exclusion process (SEP). Since then, a large amount of work has been devoted to the study of more general single-file systems 
\cite{Rodenbeck1998,Kollmann2003,Imamura2007,EuanDiaz2012,Flomenbom2008,Lizana2008,Ben-Naim2009,Manzi2012,Lizana2014,KMS2014,Barkai2009}. Most of these studies concentrate on the calculation of the variance, and even this requires a rather elaborate analysis when there are interactions in addition to the hard-core repulsion. The sub-diffusive scaling has also been verified in several experimental systems \cite{Das2010,KUKLA1996,Wei2000,Lutz2004,Lin2005,Siems2012}.

Additional interest has been triggered by the connection to interface fluctuations. The problem of tagged particle displacement can be mapped to the height fluctuation of a one-dimensional interface  \cite{Majumdar1991}. The unbiased case is related to the Edwards-Wilkinson interface growth \cite{EW1982}, whereas the biased case resembles the Kardar-Parisi-Zhang interface growth \cite{KPZ1986}.

An analytical treatment of single-file diffusion is a challenging many-body problem due to the correlations caused by the non-crossing condition. The sub-diffusive scaling of the displacement of the tagged particle, $\left \langle X_T^2 \right\rangle\sim \sqrt{T}$, is easy to understand heuristically \cite{Richards1977,Levitt1973,Alexander1978,KrapivskyBook}. An intriguing and a bit counter-intuitive property of single-file diffusion is that the variance depends on the initial state. More precisely, the scaling of the variance is robust, $\left \langle X_T^2 \right\rangle\sim \sqrt{T}$, but the pre-factor depends on whether we average only over the realizations of the stochastic dynamics or we additionally perform the averaging over initial positions \cite{ShamikSatya,Illien2013,Leibovich2013}.

Our goal is to study the higher cumulants, in fact the full statistics of the tagged particle displacement. The cumulants are encoded in the cumulant generating function 
\begin{equation}
\mu_T(\lambda)= \log \left\langle e^{\lambda X_T} \right\rangle = \lambda \left\langle X_T \right\rangle_c + \dfrac{ \lambda^2}{2!}\, \left \langle X_T^2 \right\rangle_c +  \dfrac{ \lambda^3}{3!}\,\left \langle X_T^3 \right \rangle_c + \cdots,
\label{eq:mu expansion cumulants}
\end{equation}
where $\langle X_T^n \rangle_c$ is the $n^\text{th}$ cumulant and $\lambda$  a fugacity parameter. For example, the first cumulant is  the average and the second cumulant is  the variance.

For impenetrable Brownian particles and for the SEP all the cumulants of the displacement scale as $\sqrt{T}$, see \cite{Sethuraman2013,Jara2006}. The same is expected to hold for more general single-file systems, as we shall show explicitly using a hydrodynamic formulation. In long time limit, the probability of the rescaled position of the tagged particle has a large deviation form
\begin{equation}
P\left( \dfrac{X_T}{\sqrt{4T}}=x\right)\asymp e^{-\sqrt{4T}~\phi(x)}\,.
\end{equation}
Here $\phi(x)$ is the large deviation function. The symbol $\asymp$ implies that the logarithms exhibit the same asymptotic behavior: $A\asymp B$ means that $\lim_{T\to\infty} \frac{\log A}{\log B} = 1$. The cumulant generating function $\mu_T(\lambda)$ is related to $\phi(x)$ via a Legendre transform \cite{Derrida2007,TOUCHETTE}. 

In the present work we apply the macroscopic fluctuation theory  (MFT) to single-file diffusion.
The MFT was developed by Bertini, De Sole, Gabrielli, Jona-Lasinio and Landim \cite{Bertini2001,Bertini2002,Bertini2005Current,Bertini2009,Bertini2014} for calculating large deviation functions in classical diffusive systems; similar results were obtained in the context of shot noise in conductors \cite{Sukh2004,Sukh2003}. The MFT provides a significant step towards constructing a general theoretical framework for non-equilibrium systems \cite{Derrida2007,Derrida2011}. Over the past decade the MFT has been successfully applied to numerous systems \cite{Bertini2001,Bertini2002,Bertini2005Current,Derrida2011,Bertini2006Current,Bodineau2010,Bodineau2008,Tailleur2007,Bunin2012,Gerschenfeld2009,Jona-Lasinio2010,Jona-Lasinio2014,Krapivsky2012,MS13,VMS14,Sasorov2014,MVK14,Hurtado2014,KMS_Interface,KMS2015,Yariv2015}. A perfect agreement between the MFT and microscopic calculations has been observed whenever results from both approaches were available. The MFT is a powerful and versatile tool, although the analysis is involved and challenging in most cases.

The MFT allows one to probe large deviations of macroscopic quantities such as the total current. Intriguingly, large deviations of an individual microscopic particle in a single-file system can be captured by the MFT \cite{KMS2014,KMS2015}. The essential property of single-file systems---the fixed order of the particles---allows us to express the displacement of the tagged particle in a way amenable to the MFT treatment. 

The MFT is an outgrowth of fluctuating hydrodynamic \cite{Spohn1991}. Using a path integral formulation one associates a classical action to a particular time evolution of the system. The optimal path has the least action and the analysis boils down to solving the Hamilton equations corresponding to the least-action paths. The advantage is that within this formulation all the microscopic details of the system are embedded in terms of two transport coefficients: diffusivity $D(\rho)$ and mobility $\sigma(\rho)$. These are bulk properties of the system near equilibrium, which for an arbitrary single-file system can either be measured in experiment or calculated from the microscopic dynamics.

We shall show how the cumulant generating function $\mu_T(\lambda)$, equivalently the large deviation function $\phi(x)$, can be formally expressed in terms of $D(\rho)$ and $\sigma(\rho)$. The $\sqrt{T}$ scaling of the cumulants comes out from this formal solution. The dependence of the cumulants on the initial state is also naturally incorporated within this formalism. Different initial states lead to different boundary conditions, while the governing Hamilton equations remain the same. 

Mathematically, the Hamilton equations are a pair coupled non-linear partial differential equations for two scalar fields. In the general case when $D(\rho)$ and $\sigma(\rho)$ are arbitrary, the Hamilton equations are intractable. The only solvable case corresponds to Brownian particles where $D(\rho)=1$ and $\sigma(\rho)=2\rho$. For this single-file system we deduce a closed form formula for the cumulant generating function and the associated large deviation function. We analyze annealed and quenched\footnote{We shall often use this concise, but a bit imprecise, description. A quenched state is a fixed initial state; an annealed state is actually a collection of states, say states with macroscopically uniform density, and we perform averaging over all these states.}  initial states. Our results show that the single-file system remains forever sensitive to the initial state---the large-deviation function has a very different behavior in the two cases. We verify the MFT predictions in the particular case of Brownian particles by comparing with exact results which we derive using the microscopic dynamics: the large deviation functions coming from these independent methods match perfectly. 

In the general case of arbitrary $D(\rho)$ and $\sigma(\rho)$ we use a series expansion method. In principle, the cumulants of all order can be evaluated iteratively, but in practice the calculations become very cumbersome as the order increases. Our analysis leads to an exact result for the variance of the tagged particle in terms of $D(\rho)$ and $\sigma(\rho)$. The variance in the annealed and quenched cases differ by $\sqrt{2}$. This was observed for the symmetric random average process \cite{RajeshSatya} and for
impenetrable Brownian particles \cite{Leibovich2013}, and it remains generally valid for single-file diffusion. Our general formula applies to an arbitrary single-file system and it encompasses all the results derived for specific models \cite{Harris1965,Arratia1983,Leibovich2013,Kollmann2003} as well as experimental results \cite{Lutz2004}. For the SEP, where the transport coefficients are $D(\rho)=1$ and $\sigma(\rho)=2\rho(1-\rho)$, we derive an explicit formula for the fourth cumulant. Our derivation is based on the macroscopic MFT framework, yet we also needed exact results for the integrated current which were derived in Ref.~\cite{Gerschenfeld2009Bethe} on the basis of an exact {\em microscopic} analysis employing a Bethe ansatz.

The remainder of this paper is structured as follows. In \sref{sec:mft} we start from a fluctuating hydrodynamics and show how the statistics of the tagged particle in single-file diffusion can be integrated into the MFT, leading to a variational problem. In \sref{sec:point particle}, we analyze the single-file system of Brownian point particles. A perturbative treatment of the Hamilton equations with arbitrary $D(\rho)$ and $\sigma(\rho)$ is presented in \sref{sec:series}. This allows us to derive a general formula for the variance of $X_T$. In \sref{sec:fourth}, we calculate 
the fourth cumulant for the SEP. We present a microscopic analysis of the single-file system of Brownian particles in \sref{sec:microscopic}. Some intermediate technical steps are relegated to the Appendices.

\section{A hydrodynamic formulation \label{sec:mft}}

The macroscopic fluctuation theory (MFT) is a deterministic re-formulation of fluctuating hydrodynamics. It has proven to be a very successful framework \cite{Jona-Lasinio2014,Bertini2014} for probing large deviations in diffusive particle systems (such as lattice gases). The MFT generalizes the Freidlin--Wentzell theory \cite{Freidlin1984} of finite-dimensional dynamical systems with random perturbations to a class of stochastic infinite-dimensional dynamical systems. 

The simplicity of lattice gases, and generally diffusive particle systems, is that on a macroscopic level the full description is provided by one scalar field, the density, satisfying the diffusion equation. Taking into account the stochasticity of underlying microscopic dynamics leads to a more comprehensive description known as fluctuating hydrodynamics \cite{Spohn1991}. The basic assumption is that the behavior is still essentially hydrodynamic on a large (in comparison e.g. with the lattice spacing and the hopping time) length and time scales. In one dimension,  the fluctuating hydrodynamics is based on the Langevin equation 
\begin{equation}
\partial_t\rho=\partial_x\left[ D(\rho) \partial_x\rho+\sqrt{\sigma(\rho)}~\eta \right],
\label{eq:fhe}
\end{equation}
where $\eta=\eta(x,t)$  is a Gaussian noise with covariance 
\begin{equation}
\left\langle \eta(x,t)\eta(x^{\prime},t^{\prime}) \right\rangle=\delta(x-x^{\prime})\delta(t-t^{\prime}).
\end{equation}
The diffusion coefficient $D(\rho)$ and the mobility $\sigma(\rho)$ are related to the free energy density $f(\rho)$ through the fluctuation-dissipation relation \cite{Spohn1991,Derrida2007,Krapivsky2012}
\begin{equation}
f^{\prime\prime}(\rho) = \frac{2D(\rho)}{\sigma(\rho)}
\label{eq:fluc diss}
\end{equation}
which is a consequence of the assumption of local equilibrium. 

The key feature of the fluctuating hydrodynamics and the MFT is that all microscopic details (interactions,  hopping rules, etc.) are embedded into the two transport coefficients $D(\rho)$ and $\sigma(\rho)$. These transport coefficients generically depend on the density and can be difficult to determine analytically, but they can be measured numerically or experimentally. 

The governing Langevin equation \eqref{eq:fhe} describes long time properties of the system where the density varies smoothly over the coarse graining scale. The analysis based on this equation correctly captures the large scale statistics of the fluctuations. For example, in the case of single-file diffusion only the leading time-dependence of the cumulants is correctly captured by fluctuating hydrodynamics (equivalently, the MFT).

The remarkable property that allows one to apply the MFT to single-file diffusion is a simple relation between the position of the tagged particle and the density $\rho(x,t)$ \cite{KMS2014}. Let the tagged particle starts at the origin at time $t=0$ and moves within a time window $[0,T]$. Its position $X_t$ at any time $t$ is related to the density by the single-file constraint that particles do not cross each other. This gives
\begin{equation}
\int_{X_t[\rho]}^{\infty}dx ~\rho(x,t)=\int_{0}^{\infty}dx ~\rho(x,0),
\label{eq:X_T def}
\end{equation}
thereby defining the tagged particle position\footnote{$X_t[\rho]$ is not uniquely fixed when there are regions with density equal to zero, but such configurations are highly improbable.} as a functional $X_t[\rho]$ of the density profile $\rho(x,t)$.
Equivalently, we can write 
\begin{equation}
	\int_{0}^{X_t[\rho]}dx ~\rho(x,t)=
\int_{0}^{\infty}dx \,  \big( \rho(x,t)-\rho(x,0)\big)\,.
	\label{eq:converge}
\end{equation}
In contrast to \eqref{eq:X_T def},  both integrals in \eqref{eq:converge} are convergent.

\subsection{Variational formulation}

The generating function of the cumulants of the tagged particle position $X_T$ at time $T$ can be written as a path integral of the density profile as
\begin{equation}
\left \langle e^{\lambda X_T} \right \rangle=\bigg \langle \int \mathcal{D}[\rho] ~e^{\lambda X_T[\rho]}  ~\delta \bigg(\partial_t\rho-\partial_x\big[ D(\rho) \partial_x\rho+\sqrt{\sigma(\rho)}~\eta \big] \bigg) \bigg \rangle.
\end{equation}
The Dirac delta function enforces the validity of the Langevin equation \eqref{eq:fhe}. The angular brackets denote the averaging that can consist of  two parts: over the initial density field $\rho(x,0)$ and over realizations of the stochastic noise $\eta(x,t)$ in the time window $[0,T]$ (i.e., the averaging over the history of evolution of the density profile).

If we allow the initial  state to fluctuate, we must 
 include the probability $\text{Prob}[\rho(x,0)]$ of the initial density profile. It is useful to define the  function
\begin{equation}
F[\rho(x,0)]=-\log\big(\, \text{Prob}[\rho(x,0)]\,\big).
\label{eq:F}
\end{equation}
For an initial state at equilibrium (annealed case),  $F[\rho]$ is related to the free energy. For a  quenched initial state, there are no fluctuations  of the initial profile at $t=0$ and we take the initial density profile to be the  uniform profile with density $\rho$. 

The following analysis is essentially the Martin-Siggia-Rose formalism \cite{Martin1973,DeDominicis1978}. 
Incorporating the average over the initial state, the generating function becomes
\begin{equation*}
\left \langle e^{\lambda X_T} \right \rangle=\int \mathcal{D}[\rho] ~e^{\lambda X_T[\rho]-F[\rho(x,0)]}  ~\bigg \langle \delta \bigg(\partial_t\rho-\partial_x\big[ D(\rho) \partial_x\rho+\sqrt{\sigma(\rho)}~\eta \big] \bigg) \bigg \rangle_{\eta}.
\end{equation*}
The subscript $\eta$ denotes the average over the history of noise within the window $[0,T]$ and $F$ appears only in the annealed case. Replacing the delta function by an integral over a field $\hat{\rho}$ we get
\begin{equation*}
\left \langle e^{\lambda X_T} \right \rangle =\int \mathcal{D}[\rho] ~e^{\lambda X_T[\rho]-F[\rho(x,0)]}  ~\bigg \langle \int \mathcal{D}[\hat{\rho}]~
e^{-\int_{0}^{T}dt \int_{-\infty}^{\infty} dx \hat{\rho}\left(\partial_t\rho-\partial_x\big[ D(\rho) \partial_x\rho+\sqrt{\sigma(\rho)}~\eta \big] \right) } \bigg\rangle_{\eta}.
\end{equation*}
The average with respect to the Gaussian variable $\eta(x,t)$ is computed to yield
\begin{equation*}
\left \langle e^{\lambda X_T} \right \rangle =\int \mathcal{D}[\rho,\hat{\rho}] ~e^{-S_T[\rho,\hat{\rho}]}
\label{eq:path integral}
\end{equation*}
with action 
\begin{equation}
S_T[\rho,\hat{\rho}]=-\lambda X_T[\rho]+F[\rho(x,0)]+\int_{0}^{T}dt\int_{-\infty}^{\infty} dx \bigg( \hat{\rho}\partial_t\rho-H(\rho,\hat{\rho})\bigg) 
\label{eq:action def}
\end{equation}
and Hamiltonian 
\begin{equation}
H(\rho,\hat{\rho})=\frac{\sigma(\rho)}{2}\big(\partial_x\hat{\rho}\big)^2-D(\rho)\big( \partial_x\rho\big)\big( \partial_x\hat{\rho}\big).
\label{eq:Hamiltonian}
\end{equation}
(In deriving these equations,  we  assumed that  $\hat{\rho}(x,t)\to 0$ when $x\rightarrow\pm \infty$.)

Let us rescale time by $T$ and space by $\sqrt{T}$. The tagged particle position is also rescaled by $\sqrt{T}$, so the action $S_T[\rho,\hat{\rho}]$ is proportional to $\sqrt{T}$ and we can  write  $S_T[\rho,\hat{\rho}] = \sqrt{T}S[p,q]$. 
This simple observation leads to the anomalous scaling in single-file diffusion. Indeed, for large $T$, the path integral is dominated by the least action. The cumulant generating function is then given by 
\begin{equation}
\label{TSpq}
\mu_T(\lambda)=  -  \sqrt{T} S[p,q] \, , 
\end{equation}
where we denote by $(p,q)\equiv (\hat{\rho},\rho)$ the optimal paths of the least action.\footnote{The choice of notation, $p$ and $q$, hints at the Hamiltonian nature of the governing equations \eqref{pqH}.} Equation \eqref{TSpq} implies that  at large time all the cumulants of the tagged particle position scale as $\sqrt{T}$. Note that the analysis captures only the leading $T$ dependence of the cumulants, and the sub-leading terms of the cumulants come from the correction to the saddle point approximation. 

\subsubsection*{Paths of least action}

To determine the paths of least action we consider a small variation around $(p,q)$ as $\rho=q+\delta\rho$ and $\hat{\rho}=p+\delta\hat{\rho}$. The  variation of the  action is then 
\begin{eqnarray}
\label{varS}
\delta S_T&=&\int_{0}^{T}dt \int_{-\infty}^{\infty}dx\bigg(\partial_t q-\frac{\delta H}{\delta p}\bigg)\delta\hat{\rho}(x,t)
-\int_{0}^{T}dt \int_{-\infty}^{\infty}dx\bigg(\partial_t p+\frac{\delta H}{\delta q}\bigg)\delta\rho(x,t) \nonumber\\
&+&\int_{-\infty}^{\infty}dx\bigg( -\lambda\frac{\delta X_T}{\delta q(x,0)} + \frac{\delta F}{\delta q(x,0)}-p(x,0)\bigg)\delta\rho(x,0) \nonumber\\
&+&\int_{-\infty}^{\infty}dx\bigg( -\lambda\frac{\delta X_T}{\delta q(x,T)} +p(x,T)\bigg)\delta\rho(x,T).
\end{eqnarray}
For the action to be stationary,  $\delta S_T[p,q]=0$, integrals in \eqref{varS} must vanish. Since 
$\delta\hat{\rho}(x,t)$ and $\delta\rho(x,t)$ are arbitrary, vanishing of the integrals in the first line in \eqref{varS} lead to the Hamilton equations
\begin{equation}
\label{pqH}
\partial_t q=\frac{\delta H}{\delta p}\qquad \textrm{and}\qquad \partial_t p=-\frac{\delta H}{\delta q}.
\end{equation}

The analysis of the boundary terms gives the boundary conditions for the optimal fields $(p,q)$. The density $\delta\rho(x,T)$ at final time $T$ is unconstrained, so the vanishing of the integral in the third line in \eqref{varS} leads to 
\begin{equation}
	p(x,T)=\lambda\frac{\delta X_T}{\delta q(x,T)}.
	\label{eq:pxT}
\end{equation}
The vanishing of the integral in the second line in \eqref{varS} depends on the way the system is prepared. In the  quenched case, $\delta \rho(x,0)=0$ by definition so the integral vanishes. In the annealed setting,  $\delta \rho(x,0)$ is arbitrary, therefore its pre-factor must vanish. Thus
\begin{align}
	&q(x,0)=\rho\qquad \qquad \qquad \qquad \quad  \qquad \textrm{quenched case},\\
	&p(x,0)=-\lambda\frac{\delta X_T}{\delta q(x,0)} + \frac{\delta F}{\delta q(x,0)}\qquad \textrm{annealed case}. \label{eq:px0 annealed}
\end{align}
The first equation states that the initial density in the quenched case is assumed to be uniform.

For the rest of the paper, we denote the position of the tagged particle corresponding to the least action path by
\begin{equation*}
Y\equiv X_T[q].
\end{equation*}
Note that $X_T$ is in general a random variable depending on the history of the tagged particle, but $Y$ is a deterministic quantity. 

Overall, the stochastic problem of characterizing statistics of the tagged particle position reduces to solving a variation problem. We now write the Hamilton equations and the corresponding boundary conditions by computing the functional derivatives. Different boundary conditions emerge for the quenched and annealed settings. 

\subsubsection*{Quenched case}

The Hamilton  equations read
\begin{align}
\partial_t p+D(q)\partial_{xx}p&=-\frac{\sigma^{\prime}(q)}{2}\big(\partial_x p\big)^2 \label{eq:opt p}\\
\partial_t q-\partial_x\big( D(q) \partial_x q \big)&=-\partial_x \big(\sigma(q)\partial_x p\big)
 \label{eq:opt q}
\end{align}
where $\sigma^{\prime}(q)=d\sigma(q)/dq$. 
Using (\ref{eq:X_T def}, \ref{eq:converge})  we compute the functional derivatives of $Y$ (see Appendix \ref{sec:functional derivative}) and obtain the boundary conditions
\begin{equation}
q(x,0)=\rho \qquad \textrm{and} \qquad p(x,T)= \frac{\lambda}{q(Y,T)} \, \Theta(x-Y),
\label{eq:optimal boundary quenched}
\end{equation}
where $\Theta(x)$ is the Heaviside step function.

The corresponding minimal action \eqref{eq:action def} yields the cumulant generating function 
\begin{equation*}
\mu_{\mathcal{Q}}(\lambda)=\lambda Y-\int_{0}^{T}dt\int_{-\infty}^{\infty} dx \left[p \partial_t q-\frac{\sigma(q)}{2}\big(\partial_x p\big)^2+D(q)\big( \partial_x q\big)\big( \partial_x p\big)\right].
\end{equation*}
Hereinafter we use the subscript $\mathcal{Q}$ to denote the quenched case. (For the sake of clarity, the subscript $T$ denoting the time variable is omitted from the following formulas.) The cumulant generating function can be further simplified by using Eqs.~\eqref{eq:opt p}--\eqref{eq:opt q} and integrating by parts. One gets 
\begin{equation}
\mu_{\mathcal{Q}}(\lambda)=\lambda~ Y-\int_{0}^{T}dt\int_{-\infty}^{\infty} dx\, \frac{\sigma(q)}{2}\big(\partial_x p\big)^2.
\label{eq:mu formal Quench}
\end{equation}

\subsubsection*{Annealed case}

The Hamilton  equations are the same as in the quenched case,  Eqs.~\eqref{eq:opt p}--\eqref{eq:opt q}. 
The boundary conditions are different. To derive them we use the well-known relation (see \cite{Bertini2009,Derrida2007} and \cite{Krapivsky2012} showing the consistency with the MFT and the fluctuation dissipation relation)
\begin{equation}
F[\rho(x,0)]=\int_{-\infty}^{\infty}dx\int_{\rho}^{\rho(x,0)}dr\,
\frac{2D(r)}{\sigma(r)}\, \big(\rho(x,0)-r\big).
\label{eq:F expression 1}
\end{equation}
Computing the functional derivative of  $Y$ and $F[q(x,0)]$, we recast \eqref{eq:pxT} and \eqref{eq:px0 annealed} into
\begin{align}
p(x,T)&= \frac{\lambda}{q(Y,T)} \, \Theta(x-Y), \label{eq:boundary annealed pT} \\
p(x,0)&= \frac{\lambda}{q(Y,T)}\,  \Theta(x)+\int_{\rho}^{q(x,0)}dr\,\frac{2D(r)}{\sigma(r)}\,.  \label{eq:boundary annealed p0}
\end{align}
The least action \eqref{eq:action def}
can be simplified using \eqref{eq:F expression 1} and the Hamilton  equations.
The cumulant  generating function for the annealed initial state becomes 
\begin{equation}
\mu_{\mathcal{A}}(\lambda)=\lambda ~Y-\int_{-\infty}^{\infty}dx\int_{\rho}^{q(x,0)}dr
\frac{2D(r)}{\sigma(r)} \bigg( q(x,0)-r \bigg) -\int_{0}^{T}dt\int_{-\infty}^{\infty} dx \frac{\sigma(q)}{2}\big(\partial_x p\big)^2.
\label{eq:mu formal annealed}
\end{equation}
Hereinafter the subscript ${\mathcal{A}}$ denotes the annealed case.

\medskip
\noindent 
{\it A Remark on Symmetry.} It is instructive to look at the symmetry properties of the action and optimal fields. One can notice symmetry relations 
\begin{align}
q_{-\lambda}(-x,t) &= q_{\lambda}(x,t) \label{eq:symmetry q}\\
p_{-\lambda}(-x,t) &= p_{\lambda}(x,t)-\dfrac{\lambda}{q_{\lambda}(Y_{\lambda},T)}  \label{eq:symmetry p}\\
Y_{-\lambda} &= -Y_{\lambda}  \label{eq:symmetry Y}\\
\mu_T(-\lambda) &=\mu_T(\lambda) \label{eq:symmetry mu}
\end{align}
underlying  the variational problem in both quenched and annealed settings. Since $\mu(\lambda)$ is an even function, 
all odd cumulants of the tagged particle position are zero. This is the consequence of the fact that the microscopic dynamics is unbiased.

\section{Brownian particles with hard-core repulsion \label{sec:point particle}}

\begin{figure}
\begin{center}
\includegraphics[scale=0.75]{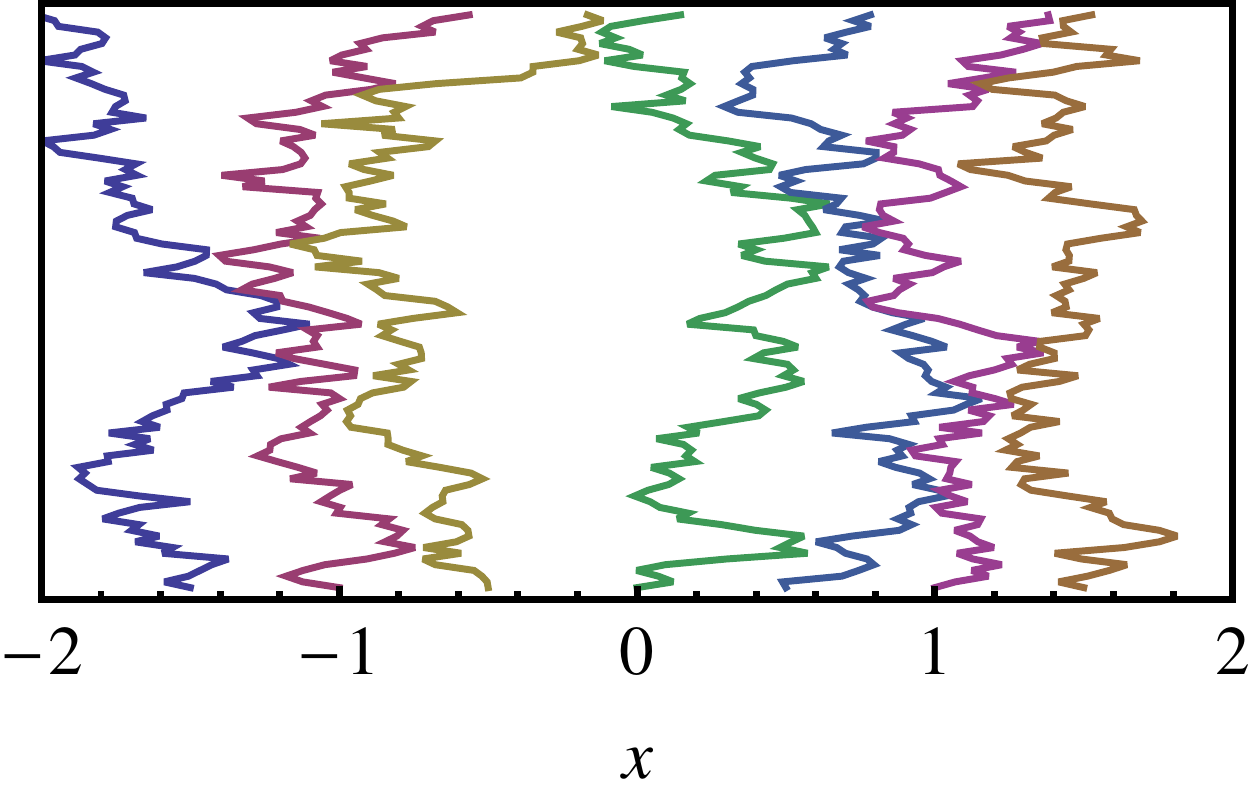}
\end{center}
\caption{A sample trajectory of Brownian point particles with hard-core repulsion. The trajectories may come infinitely close but never cross each other, keeping the order of particles unchanged.  \label{fig:trajectory}}
\end{figure}

It is hard to analyze the variational problem for a general single-file system characterized by density-dependent transport coefficients $D(\rho)$ and $\sigma(\rho)$. The only system that is amenable to analyses, both macroscopic (based on the MFT) and microscopic, is the system of Brownian point particles with hard-core repulsion (Fig.~\ref{fig:trajectory}). The system was first studied by Harris \cite{Harris1965} who used a mapping to non-interacting particles to compute the variance of the tagged particle position. The problem was later studied by many other authors  
 (see e.g. \cite{Leibovich2013,Hegde2014,Lizana2008,Rodenbeck1998}). 

For Brownian particles the diffusion coefficient is constant, $D(\rho)=1$, and the mobility is a linear function of density, $\sigma(\rho)=2\rho$. The Hamilton equations \eqref{eq:opt p}--\eqref{eq:opt q} become
\begin{align}
\partial_t p+\partial_{xx}p &= -\big( \partial_x p \big)^2, \label{eq:opt p point}\\
\partial_t q-\partial_{xx}q&= -\partial_x \big(2q \partial_x p \big). \label{eq:opt q point}
\end{align}
A canonical  Hopf-Cole transformation,  $(p,q)\rightarrow(P,Q)=\left(e^p, q e^{-p}\right)$, is known \cite{Gerschenfeld2009,Sasorov2014,Elgart2004} to simplify Eqs.~\eqref{eq:opt p point}--\eqref{eq:opt q point}. The new conjugate variables satisfy the Hamilton  equations with $H=-(\partial_x P)(\partial_x Q)$. In the $(P,Q)$ variables, the governing equations are (linear) anti-diffusion and diffusion equations:
\begin{equation}
\partial_t P+\partial_{xx}P=0 \quad \textrm{and}\quad \partial_t Q -\partial_{xx}Q=0.
\end{equation}
Solving these equations and returning to the original variables $(p,q)$ we arrive at a formal solution
\begin{align}
p(x,t)&= \log \left( \int_{-\infty}^{\infty}dz~e^{p(z,T)} ~\frac{\exp\left(-\frac{(z-x)^2}{4 (T-t)} \right)}{\sqrt{4\pi (T-t)}}\right),\label{eq:p gen sol point}\\
q(x,t)&= \int_{-\infty}^{\infty}dz~q(z,0)~e^{p(x,t)-p(z,0)}~\frac{\exp\left(-\frac{(z-x)^2}{4 t} \right)}{\sqrt{4\pi t}}
 \label{eq:q gen sol point}
\end{align}
applicable to both quenched and annealed settings and to arbitrary $p(x,T)$ and $q(x,0)$. 

\subsection{Quenched case}

Let us re-write \eqref{eq:optimal boundary quenched} as 
\begin{equation}
q(x,0)=\rho \qquad \textrm{and} \qquad p(x,T)=B~\Theta(x-Y).
\label{eq:boundary quenched}
\end{equation}
Here we have used a shorthand notation
\begin{equation}
B=\frac{\lambda}{q(Y,T)}\,.
\label{eq:B def}
\end{equation}
The quantity $Y$ is related to $q(x,t)$ via  \eqref{eq:converge}  which becomes 
\begin{equation}
Y=\int_{Y}^{\infty}dx \left( \frac{q(x,T)}{\rho}-1\right).
\label{eq:Y def quench}
\end{equation}
in the quenched case. 

The boundary conditions \eqref{eq:boundary quenched} depend on the solution itself. Let us proceed by treating  $B$ and $Y$ as parameters to be determined later. The optimal fields in terms of $B$ are 
\begin{align}
p(x,t)&= \log\left[\, 1+\left( e^B-1 \right) \frac{1}{2} \erfc{\frac{Y-x}{\sqrt{4(T-t)}}} \, \right], \label{eq:p quench final point}\\
q(x,t)&= \rho \int_{-\infty}^{\infty}dz~\left(\frac{1+\left( e^B-1 \right) \frac{1}{2} \erfc{\frac{Y-x}{\sqrt{4(T-t)}}} }{1+\left( e^B-1 \right) \frac{1}{2} \erfc{\frac{Y-z}{\sqrt{4T}}} }\right)~\frac{\exp\left(-\frac{(z-x)^2}{4 t} \right)}{\sqrt{4\pi t}}, \label{eq:q quench final point}
\end{align}
where $\erfc{x}$ is the complementary error function. Substituting $D(\rho)=1$ and $\sigma(\rho)=2\rho$ into the formula for the cumulant generating function  \eqref{eq:mu formal Quench} leads to
\begin{equation*}
\mu_{\mathcal{Q}}(\lambda)=\lambda~Y-\int_{0}^{T}dt\int_{-\infty}^{\infty}dx~q(\partial_x p)^2.
\end{equation*}
 This expression for $\mu_{\mathcal{Q}}(\lambda)$ can be simplified thanks to identity
\begin{equation}
q(\partial_x p)^2=\partial_t(q~p)-\partial_x \left( p\partial_xq-q\partial_xp -2qp\partial_xp \right),
\label{identite:utile}
\end{equation}
which results  from Eqs.~\eqref{eq:opt p point}--\eqref{eq:opt q point}. Using \eqref{identite:utile} and $p(x,t)\rightarrow 0$ as $x\rightarrow \pm \infty$,  we obtain 
\begin{equation*}
\mu_{\mathcal{Q}}(\lambda)=\lambda~Y-\int_{-\infty}^{\infty}dx~q(x,T)p(x,T)
+\int_{-\infty}^{\infty}dx~q(x,0)p(x,0).
\end{equation*}
Plugging into this formula $p(x,T)$ and $q(x,0)$ from \eqref{eq:boundary quenched} we get
\begin{equation*}
\mu_{\mathcal{Q}}(\lambda)=\lambda~Y-B\int_{Y}^{\infty}dx~q(x,T)
+\rho\int_{-\infty}^{\infty}dx~p(x,0).
\end{equation*}
It is clear from Eqs.~\eqref{eq:p quench final point}--\eqref{eq:q quench final point} that the functions $q(x,T)$ and $p(x,0)$ approach to non-zero constants as $x\rightarrow\infty$. This implies that the integrals in the above formula are not convergent although their linear combination is well defined. To write $\mu_{\mathcal{Q}}(\lambda)$ in terms of convergent integrals we use the definition of $Y$ in \eqref{eq:Y def quench} and subsequently rearrange the integrals and obtain 
\begin{equation*}
\mu_{\mathcal{Q}}(\lambda)=(\lambda-\rho B)~Y+\rho\int_{Y}^{\infty}dx~\big( p(x,0)-B \big)
+\rho\int_{-Y}^{\infty}dx~p(-x,0).
\end{equation*}
The integrals are  now written in terms of $p(x,0)$ and,  using \eqref{eq:p quench final point},  the formula for the cumulant generating function becomes
\begin{align*}
\frac{\mu_{\mathcal{Q}}(\lambda)}{\sqrt{4T}} = (\lambda-\rho B)\,y 
&+\rho\int_{0}^{\infty}d\xi~ \log\left(1+\frac{e^{-B}-1}{2}\, \erfc{\xi} \right)\\
& +\rho\int_{0}^{\infty}
d\xi~ \log\left(1+\frac{e^{B}-1}{2}\, \erfc{\xi} \right).
\end{align*}
where $y=Y/\sqrt{4T}$ and $\xi=x/\sqrt{4T}$.  
This is further simplified by using $\erfc{\xi}+\erfc{-\xi}=2$ to give
\begin{equation}
\frac{\mu_{\mathcal{Q}}(\lambda)}{\sqrt{4T}} = (\lambda-\rho B)\,y
+\rho\int_{0}^{\infty}d\xi~\log\left( 1+\sinh^2\left( \frac{B}{2} \right)\erfc{\xi} \erfc{-\xi}\right).
\label{eq:mu Q 1}
\end{equation}
The reason for writing the formula in this form will become clear shortly. 

So far, $Y$ and $B$ have been treated as parameters. One relation between these parameters is obtained by inserting \eqref{eq:q quench final point} into \eqref{eq:Y def quench}:
\begin{equation*}
y=\frac{e^B-1}{2} \int_{-\infty}^{\infty}d\xi~
\frac{\erfc{\xi-y}\erfc{y-\xi}}{2+\left(e^B-1\right)\erfc{y-\xi}}.
\end{equation*}
Changing the variable $\xi\rightarrow \xi+y$ and using $\erfc{\xi}+\erfc{-\xi}=2$ we transform the above formula into
\begin{align*}
y= \frac{e^B-1}{2} \int_0^{\infty}d\xi~
\frac{\erfc{\xi}\erfc{-\xi}}{2+\left(e^B-1\right)\erfc{\xi}}
- \frac{e^{-B}-1}{2} \int_0^{\infty}d\xi~
\frac{\erfc{\xi}\erfc{-\xi}}{2+\left(e^{-B}-1\right)\erfc{\xi}}
\end{align*}
Massaging this formula one arrives at a more neat form
\begin{equation}
y=\frac{d}{dB}\int_{0}^{\infty}d\xi~\log\left( 1+\sinh^2\left( \frac{B}{2} \right)\erfc{\xi} \erfc{-\xi}\right).
\label{eq:YT Q 1}
\end{equation}

\begin{figure}
\begin{center}
\includegraphics[scale=1]{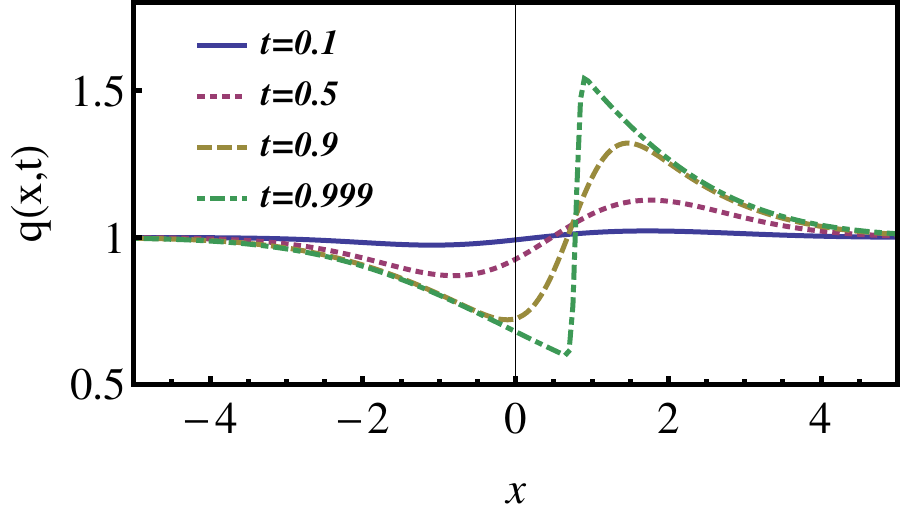}
\end{center}
\caption{The optimum density field $q(x,t)$, Eq.~\eqref{eq:q quench final point}, at different times. We set $\rho=1$, $B=2$ and $T=1$, and determined $Y$ from \eqref{eq:YT Q 1}. The density at initial time $t=0$ starts from the quenched uniform profile $q(x,0)=1$ and as time approaches $t=T$, the profile develops a sharp jump at the position of the tagged particle $Y\simeq 1.54$, indicating a discontinuity of the function $q(x,T)$ at $x=Y$.  \label{fig:singularity}}
\end{figure}

A similar self-consistent way of determining $B$ using relation \eqref{eq:B def} does not lead to a unique value for $B$. This is because the solution $q(x,t)$ in \eqref{eq:q quench final point} is singular at $(x,t)\equiv (Y,T)$. A graphical representation of this singularity is shown in \fref{fig:singularity}. The problem is analogous to the case of a diffusion equation with a step initial profile: The solution at any time $t>0$ is independent of the precise value of the initial profile at the position of the step.

For any value of $B$, Eqs.~\eqref{eq:p quench final point}--\eqref{eq:q quench final point} give a solution of the Hamilton  equations. Only one solution corresponds to the minimum action.  This  solution 
can be determined by optimizing the action with respect to $B$, i.e., by imposing 
\begin{equation}
\frac{d\mu_{\mathcal{Q}}(\lambda)}{dB}=0\,.
\end{equation}
Combining this with \eqref{eq:mu Q 1} and \eqref{eq:YT Q 1} we find the optimal $B$:
\begin{equation}
B=\frac{\lambda}{\rho}\,.
\label{eq:B}
\end{equation}
Putting above in \eqref{eq:mu Q 1} we arrive at an explicit formula for the cumulant generating function:
\begin{equation}
\frac{\mu_{\mathcal{Q}}(\lambda)}{\sqrt{4T}}=\rho\int_{0}^{\infty}d\xi~ \log\left(1+\sinh^2\left( \frac{\lambda}{2\rho} \right)\erfc{\xi} \erfc{-\xi} \right).
\label{eq:mu Q final}
\end{equation}

\subsubsection*{The large deviation function}
The large deviation function is related to $\mu_{\mathcal{Q}}(\lambda)$ via the Legendre transform: 
$\phi_{\mathcal{Q}}(y)= \sup_\lambda \left(\lambda y-\frac{\mu_{\mathcal{Q}}(\lambda)}{\sqrt{4T}} \right)$. Using \eqref{eq:mu Q final} and \eqref{eq:YT Q 1} we can represent the large deviation function in the parametric form
\begin{equation}
\rho^{-1} \phi_{\mathcal{Q}}(y)=By
-\int_{0}^{\infty}d\xi~ \log\left(1+\sinh^2\!\left( \frac{B}{2} \right)\erfc{\xi} \erfc{-\xi} \right)
\label{eq:phi Q final}
\end{equation}
with  $B$ determined from the optimality requirement
\begin{equation}
\frac{d\phi_{\mathcal{Q}}}{dB}=0.
\label{eq:y Q final}
\end{equation}
An equivalent representation of $\phi_{\mathcal{Q}}(y)$ is
\begin{align}
\rho^{-1}~\phi_{\mathcal{Q}}(y)= - \int_{-y}^{\infty}d\xi~ \log\left(1+\frac{e^{-B}-1}{2}\, \erfc{\xi} \right)
-\int_{y}^{\infty}
d\xi~ \log\left(1+\frac{e^{B}-1}{2}\, \erfc{\xi} \right).
\label{eq:ldf Q alternate form}
\end{align}

All previous results have been derived using a macroscopic approach. In \sref{sec:microscopic} we show that the same expression for $\phi_{\mathcal{Q}}$ follows from an exact microscopic analysis. This is reassuring since the macroscopic approach is not fully rigorous, yet much more widely applicable than exact analyses which are limited to simplest systems.

\subsection{Annealed case}

The major difference with the quenched case comes from the boundary conditions \eqref{eq:boundary annealed p0}--\eqref{eq:boundary annealed pT}. When $D(\rho)=1$ and $\sigma(\rho)=2\rho$, the boundary conditions become
\begin{align}
q(x,0) &=  \rho~\exp\big( p(x,0)-B~\Theta(x) \big) \label{eq:boundary point q ann},\\
p(x,T) &= B~\Theta(x-Y).  \label{eq:boundary point p ann}
\end{align}
The parameter $B$ is again defined in \eqref{eq:B def} and $Y$ is determined from \eqref{eq:converge}, equivalently
\begin{equation}
	\int_{0}^{Y}dx ~q(x,T)=\int_{0}^{\infty}dx~\big( q(x,T)-q(x,0) \big).
	\label{eq:Yconverge}
\end{equation}
The boundary condition depends on the solution itself and similar to the quenched case, a solution of the optimal fields is found by treating $Y$ and $B$ as parameters. They are later determined using the solution. Substituting the boundary condition in the general solution \eqref{eq:p gen sol point}--\eqref{eq:q gen sol point} leads to a formula of the optimal fields in terms of the $Y$ and $B$.
\begin{align}
p(x,t)&= \log\left( \, 1+\left( e^B-1 \right) \frac{1}{2} \erfc{\frac{Y-x}{\sqrt{4(T-t)}}}\,  \right), \label{eq:p annealed point final}\\
\frac{q(x,t)}{\rho}&= \left(1+\left( e^{-B}-1 \right) \frac{1}{2} \erfc{\frac{x-Y}{\sqrt{4(T-t)}}} \right) \Bigg(1+\left( e^{B}-1 \right) \frac{1}{2} \erfc{\frac{x}{\sqrt{4t}}} \Bigg).
 \label{eq:q annealed point final}
\end{align}
Equation \eqref{eq:mu formal annealed} 
for the cumulant generating function becomes 
\begin{equation*}
\mu_{\mathcal{A}}(\lambda)=\lambda ~Y -\int_{-\infty}^{\infty}dx\int_{\rho}^{q(x,0)}dr
 \bigg( \frac{q(x,0)}{r}-1 \bigg)-\int_{0}^{T}dt\int_{-\infty}^{\infty} dx~ q\big(\partial_x p\big)^2,
\end{equation*}
where we have used again $D(\rho)=1$ and $\sigma(\rho)=2\rho$. This  formula  for $\mu_{\mathcal{A}}(\lambda)$ can be rewritten, thanks to the identity \eqref{identite:utile},  as 
\begin{align*}
\mu_{\mathcal{A}}(\lambda)= \lambda~ Y &-\int_{-\infty}^{\infty}dx~q(x,0)\log\left(  \frac{q(x,0)}{\rho}\right)+\int_{-\infty}^{\infty}dx \big(q(x,0)-\rho  \big)\\
&-\int_{-\infty}^{\infty}dx~q(x,T)p(x,T)
+\int_{-\infty}^{\infty}dx~q(x,0)p(x,0),
\end{align*}
(we have also taken into account that $p(x,t)=0$ at $x\rightarrow\pm\infty$ at all time $t$). Using \eqref{eq:boundary point q ann}--\eqref{eq:Yconverge} one can greatly simplify the above expression:
\begin{equation*}
\mu_{\mathcal{A}}(\lambda)= \lambda~ Y+\int_{-\infty}^{\infty}dx~\big(q(x,0)-\rho \big).
\end{equation*}
Combining this with \eqref{eq:q annealed point final} we find 
\begin{align}
\frac{\mu_\mathcal{A}(\lambda)}{\sqrt{4T}} = \lambda y  +\rho\, \frac{e^B -1}{2}\int_{y}^{\infty}d\xi~\erfc{\xi}
+\rho\, \frac{e^{-B} -1}{2}\int_{-y}^{\infty}d\xi~\erfc{\xi}.
\label{eq:mu Ann final}
\end{align}

So far, $Y$ and $B$ were treated as parameters. Using \eqref{eq:q annealed point final} one can recast \eqref{eq:Yconverge} into 
\begin{equation}
2y= \left( e^B -1 \right)\int_{y}^{\infty}d\xi~\erfc{\xi}
- \left( e^{-B} -1 \right)\int_{-y}^{\infty}d\xi~\erfc{\xi}.
\label{eq:Y annealed point final}
\end{equation}

\begin{figure}
\begin{center}
\includegraphics[scale=1]{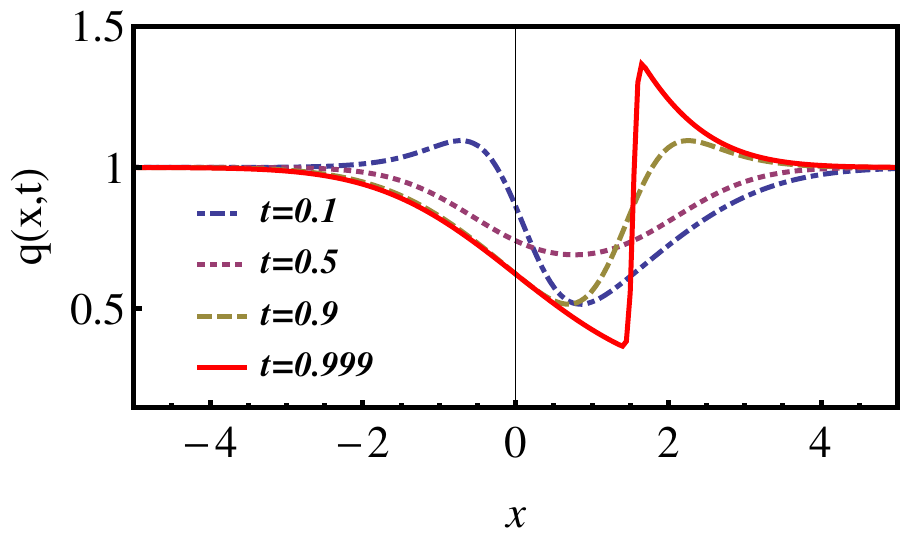}
\end{center}
\caption{The optimum density field $q(x,t)$ in \eqref{eq:q annealed point final} for the annealed case at different times. We set $Y=1.5$ and $T=1$ and determined $B$ from Eq.~\eqref{eq:Y annealed point final}. As time approaches $t=T=1$, the profile develops a sharp jump at the position $Y$ of the tagged particle, indicating a discontinuity of the function $q(x,T)$. Note that the profile $q(x,t)$ is symmetric under $(x,t)\rightarrow(Y-x,T-t)$ as can be seen from Eq.~\eqref{eq:q annealed point final}. \label{fig:singularityAnnealed}}
\end{figure}

The  second parameter $B$ cannot be obtained by evaluating  $q(x,T)$ at $x=Y$ because this function
 is singular. (This singularity is evident from \fref{fig:singularityAnnealed}.) As in the quenched case, the parameter $B$ has to be determined from an optimization criterion $\frac{d \mu_{\mathcal{A}}}{dB}=0$. Using this together with \eqref{eq:mu Ann final} we derive a relation between  $B$, $Y$ and  the fugacity parameter $\lambda$: 
\begin{equation}
\frac{2\lambda}{\rho}= \left( e^B -1 \right) \erfc{y}
- \left( e^{-B} -1 \right)  \erfc{-y}.
\label{eq:lambda annealed point final}
\end{equation}
To derive \eqref{eq:lambda annealed point final} we used 
\begin{equation}
\label{Id:2}
e^{B}\int_{y}^{\infty}d\xi~\erfc{\xi} = e^{-B} \int_{-y}^{\infty}d\xi~\erfc{\xi}
\end{equation}
which follows from \eqref{eq:Y annealed point final} in conjunction with the  elementary identity 
\begin{equation}
\label{Id:1}
\int_{-y}^{\infty}d\xi~\erfc{\xi} - \int_{y}^{\infty}d\xi~\erfc{\xi} = 2y.
\end{equation}
Equations \eqref{eq:mu Ann final}--\eqref{eq:lambda annealed point final} constitute a parametric solution for the cumulant generating function in the annealed setup. 
Unlike the quenched case, we don't have an explicit formula for $\mu_{\mathcal{A}}(\lambda)$.  
Another parametric representation
\begin{align*}
\frac{\mu_{\mathcal{A}}(\lambda)}{\rho\sqrt{4T}}&= \left(\frac{\lambda}{\rho}  +
 \frac{1 - e^B}{1 +  e^B}  \right) y , \\
  e^{2B}  & = 1  +  2y \, \left[ \int_{y}^{\infty}du~\erfc{u} \right]^{-1}, \\
\frac{\lambda}{\rho}& = \left(1 -  e^{-B} \right) \left(1 + \frac{ e^B -1 }{2}\, \erfc{y} \right)
\end{align*}
has been reported in Refs.~\cite{Hegde2014,KMS2014}. It
can be obtained  from \eqref{eq:mu Ann final}--\eqref{eq:lambda annealed point final} using \eqref{Id:2} and \eqref{Id:1}.

\subsubsection*{The large deviation function}

The large deviation function is again the Legendre transform: 
$\phi_{\mathcal{A}}(y)= \sup_\lambda \left(\lambda y-\frac{\mu_{\mathcal{A}}(\lambda)}{\sqrt{4T}} \right)$. Combining it with \eqref{eq:mu Ann final}
we obtain 
\begin{equation*}
\frac{2\phi_{\mathcal{A}}(y)}{\rho}=-\left( e^B -1 \right)\int_{y}^{\infty}d\xi\,\erfc{\xi}
- \left( e^{-B} -1 \right)\int_{-y}^{\infty}d\xi\,\erfc{\xi}.
\end{equation*}
Using Eq.~\eqref{Id:2} we eliminate the dependence on $B$ and arrive at the following explicit formula for the large deviation function
\begin{equation}
\frac{2\phi_{\mathcal{A}}(y)}{\rho}=\left\{ \sqrt{\int_{y}^{\infty}d\xi\,\erfc{\xi}}-  \sqrt{\int_{-y}^{\infty}d\xi\,\erfc{\xi}} \right\}^2.
\label{eq:ldf final}
\end{equation}
This  result will be re-derived in \sref{sec:microscopic} using an exact microscopic  analysis.

\subsection{Comparing  annealed and quenched settings}
\subsubsection*{The cumulants}

By definition \eqref{eq:mu expansion cumulants}, the cumulants are obtained by  expanding the cumulant generating function. For the quenched initial condition this expansion of $\mu_{\mathcal{Q}}(\lambda)$ is simple to generate from the explicit formula \eqref{eq:mu Q final}. We write below the first few cumulants of the tagged particle position,
\begin{align}
		\left\langle X_T^{2} \right \rangle_c &=  \frac{1}{\rho}\, I_{1}\sqrt{T}, \label{eq:X2 quenched point}\\
		\left\langle X_T^{4} \right \rangle_c & = \frac{1}{\rho^{3}}
\left(I_{1}-\frac{3}{2} I_{2}\right)\sqrt{T},\\
		\left\langle X_T^{6} \right\rangle_c& =\frac{1}{\rho^{5}}\left[I_{1}-\frac{15}{2}\big(
		I_{2}- I_{3}\big)\right]\sqrt{T},
\end{align}
where $I_{n}=\int_{0}^{\infty}dz\,[\erfc{z}\erfc{-z}]^n$. The first two integrals $I_1$ and $I_2$ are known \cite{Prudnikov1986} leading to 
\begin{align}
		\left\langle X_T^{2} \right \rangle_c &=  
		\frac{\sqrt{2}}{\rho\sqrt{\pi}}\sqrt{T},\\
		\left\langle X_T^{4} \right \rangle_c &=  \frac{1}{\rho^{3}}\left(
		\frac{9}{\pi}\arctan\left( \frac{1}{2\sqrt{2}} \right)-1
		\right)\frac{2\sqrt{2}}{\sqrt{\pi}}\sqrt{T}.
\label{cumulant4:Bquenched}
\end{align}

For the annealed case, the generating function has a parametric form. To make a series expansion in powers of $\lambda$, we first use Eq.~\eqref{eq:Y annealed point final} to eliminate $B$ from \eqref{eq:mu Ann final} and subsequently make an expansion of the $\mu_{\mathcal{A}}(\lambda)$ in powers of $y$. Then we use \eqref{eq:Y annealed point final} and \eqref{eq:lambda annealed point final} to make an expansion of $y$ in terms of $\lambda$ leading to an expansion of $\mu_{\mathcal{A}}(\lambda)$ in powers of $\lambda$. The first three non-trivial cumulants are:
		\begin{align}
		\left\langle X_T^{2} \right \rangle_c &=  \frac{2}{\rho\sqrt{\pi}}\, \sqrt{T}, \label{eq:X2 annealed point}\\
		\left\langle X_T^{4} \right \rangle_c &=  \frac{1}{\rho^{3}}\left(\frac{4}{\pi}-1\right)\frac{6}{\sqrt{\pi}}\sqrt{T}, \label{eq:X4 annealed point}\\
		\left\langle X_T^{6} \right\rangle_c &= \frac{1}{\rho^{5}}\left(\frac{408}{\pi^2}-\frac{180}{\pi}+18\right)\frac{5}{\sqrt{\pi}}\sqrt{T}. \label{eq:X6 anneal point}
	\end{align}
	
\begin{figure}[t]
	\begin{center}
	\includegraphics[width=0.6\textwidth]{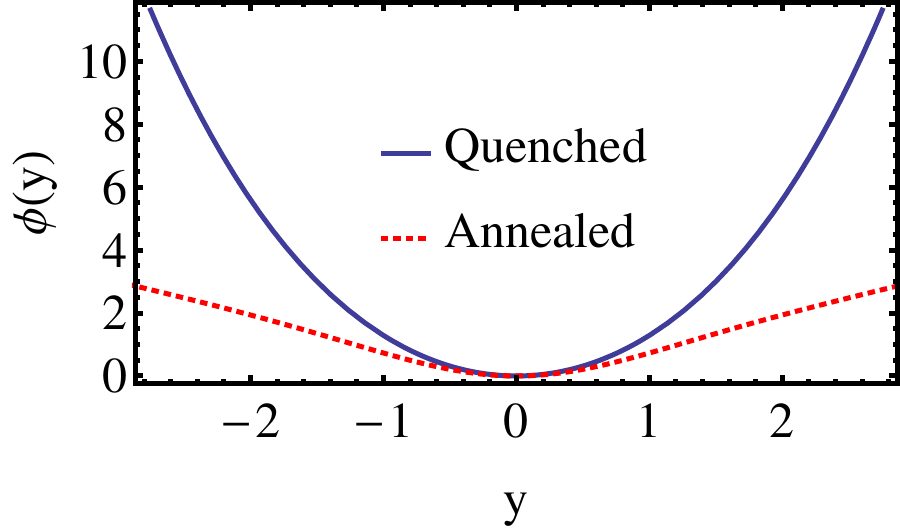}
\caption{The large deviation function $\phi(y)$ of the tagged particle position in the annealed and quenched settings. In both cases, the initial average density is chosen to be uniform $\rho=1$. Note that the quenched large deviation function is always larger, see Appendix  \ref{app:inequality} for a theoretical explanation. \label{fig:ldf}}
	\end{center}
\end{figure}

\begin{figure}[t]
	\begin{center}
	\includegraphics[width=0.6\textwidth]{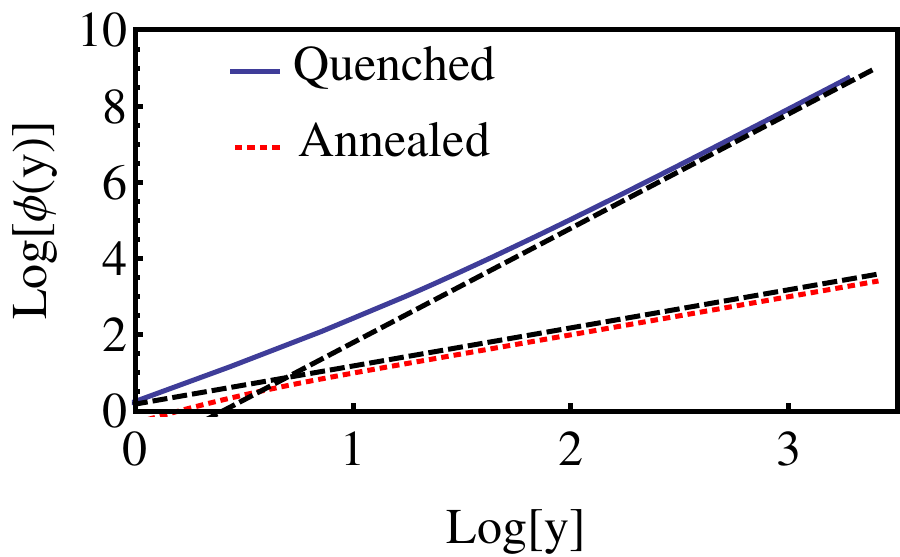}
	\caption{Asymptotics of the large deviation functions for large values of $y$. The dashed straight lines correspond to the power laws $y^3$ and $y$, for the quenched and the annealed case, respectively. \label{fig:ldfasymptotic}}
	\end{center}
\end{figure}

\subsubsection*{The large deviation function}

The large deviation function depends on the setting. Strikingly different asymptotic behaviors emerge in the annealed and quenched settings (Figs.~\ref{fig:ldf} and \ref{fig:ldfasymptotic}). In both cases, the large deviation function has a non-Gaussian tail. More precisely, the large deviation function $\phi(y)$ grows linearly in the annealed case and cubically in the quenched case (see Fig.~\ref{fig:ldfasymptotic}):
\begin{equation}
\label{FF}
\phi_{\mathcal{A}}(y) \simeq  \rho ~|y|, \qquad 
\phi_{\mathcal{Q}}(y) \simeq  \frac{\rho}{3}~|y|^3.
\end{equation}
In the annealed case, the large $y$ is derived from \eqref{eq:ldf final} and the asymptotic relations:
\begin{align*}
\int_{y}^{\infty}dz \frac{1}{2}\erfc{z}&=e^{-y^2}\bigg(\frac{1}{4\sqrt{\pi}~y^2}+\mathcal{O}(y^{-4})\bigg),\\
\int_{-y}^{\infty}dz \frac{1}{2}\erfc{z}&= y+e^{-y^2}\bigg(\frac{1}{4\sqrt{\pi}~y^2}+\mathcal{O}(y^{-4})\bigg).
\end{align*}
In the quenched case, the large deviation function has a parametric form, making the analysis a bit more involved. We use \eqref{eq:phi Q final} and \eqref{eq:y Q final} and notice that large $y$ corresponds to large $B$. The integral in \eqref{eq:phi Q final} grows as $\frac{2}{3}B^{3/2}$ for large $B$. Indeed, we divide the range of integration, $\int_{0}^{\infty}dx=\int_{0}^{\sqrt{B}}dx+\int_{\sqrt{B}}^{\infty}dx$, use the $x\to\infty$ asymptotic 
\begin{equation*}
	\erfc{x}\erfc{-x} = e^{-x^2}\left( \dfrac{2}{\sqrt{\pi}~x}+\mathcal{O}\left( \dfrac{1}{x^3}\right) \right),
\end{equation*}
and write
\begin{equation*}
\int_{0}^{\sqrt{B}}dx\,\log\left(1+\sinh^2\left( \frac{B}{2} \right) \erfc{x} \erfc{-x} \right)=\int_{0}^{\sqrt{B}}dx~\big(B-x^2\big)+\mathcal{O}(\sqrt{B}),
\end{equation*}
and find $\frac{2}{3}B^{3/2}$. This is the dominant contribution. The second integral contributes terms of order $1/B$ only. Thus $\rho^{-1} \phi_{\mathcal{Q}}(y)\simeq By-\frac{2}{3}B^{3/2}$ which in conjunction with \eqref{eq:y Q final} leads to the asymptotic in \eqref{FF}.

A similar asymptotic dependence is also observed in the large deviation function for the current in the symmetric exclusion process with step initial condition \cite{Gerschenfeld2009,Sasorov2014}.
It would be interesting to study how universal are these  power-law
tails. An analogy between the displacement of the tagged particle and
current suggests that the tails are non-universal, as for the
extreme current fluctuations \cite{MS13,VMS14}.

\section{Calculation of the variance for general single-file systems \label{sec:series}}

For a single-file system with arbitrary $D(\rho)$ and $\sigma(\rho)$ it is impossible to solve the Hamilton equations. One can try to seek an asymptotic solution \cite{Krapivsky2012} as a series expansion in powers of $\lambda$. This series solution is then used to expand the cumulant generating function. 

The expansions for the optimal fields read
\begin{align}
	q(x,t)&=\rho+\lambda ~q_{1}(x,t)+\lambda^{2}~q_{2}(x,t)+\cdots,\label{eq:expansion 1}\\
	p(x,t)&=\lambda~ p_{1}(x,t)+\lambda^{2}~p_{2}(x,t)+\cdots,
	\label{eq:expansion}
\end{align}
In the zeroth order, the solution is deterministic: $q(x,t)=\rho$ and $p(x,t)=0$. This follows from the governing Eqs.~\eqref{eq:opt p}--\eqref{eq:opt q} both in the quenched and annealed cases. In the first order 
\begin{eqnarray}
	\partial_{t}p_{1}+D(\rho)\partial_{xx}p_{1}&&= 0, \label{eq:optimal p1}\\
	\partial_{t}q_{1}-D(\rho)\partial_{xx}q_{1}&&=-\sigma(\rho)\partial_{xx}p_{1}.
	\label{eq:optimal q1}
\end{eqnarray}
Using \eqref{eq:mu formal Quench} and \eqref{eq:mu formal annealed} we see that the variance
\begin{equation}
\frac{1}{2!}\left \langle X_{T}^{2} \right\rangle= Y_{1}-F_{2}-\frac{\sigma(\rho)}{2}\int_{0}^{T}dt\int_{-\infty}^{\infty}dx
	\left( \partial_{x}p_{1}\right)^{2}
	\label{eq:X2}
\end{equation}
explicitly depends on $p_1, Y_1,F_2$; it also on $q_1$ (through $Y_1, F_2$). Since $Y$ is an odd function of $\lambda$, see  \eqref{eq:symmetry Y}, the zeroth order term vanishes while the first order term is given by
\begin{equation}
	Y_{1}=\frac{1}{\rho}\int_{0}^{\infty}dx \left(q_{1}(x,T)-q_{1}(x,0)
	\right)
\label{eq:Y1}
\end{equation}
as it follows from (\ref{eq:X_T def},\ref{eq:converge}). It remains to determine $F_{2}$. In the quenched case, $F$ does not appear (one can formally set $F \equiv 0$ in the quenched setting). In the annealed case we use \eqref{eq:F expression 1} and find
\begin{equation}
	F_{2}=\frac{D(\rho)}{\sigma(\rho)}\int_{-\infty}^{\infty}dx\left( q_1(x,0)
	\right)^2 .
	\label{eq:F2 annealed}
\end{equation}

We now determine the variance $\langle X_{T}^{2} \rangle$ by
solving \eqref{eq:optimal p1}--\eqref{eq:optimal q1} with the
specific boundary conditions for the quenched and the annealed initial states.

\subsection{The quenched initial state}

Plugging the series expansion into the boundary condition \eqref{eq:optimal boundary quenched} yields
\begin{equation}
	p_{1}(x,T)=\rho^{-1}\Theta(x)\qquad \textrm{and}\qquad
	q_{1}(x,0)=0
\label{eq:boundary linear quenched}
\end{equation}	
in the linear order. With the above boundary condition
the solution of \eqref{eq:optimal p1} can be expressed as 
\begin{equation}
	\partial_{x}p_{1}(x,t)=\rho^{-1}~g(0,T \vert x,t),
	\label{eq:dp1 quench}
\end{equation}
where $g$ is the diffusion propagator 
\begin{equation}
\label{g:propagator}
g(z,T \vert x, t)=\frac{1}{\sqrt{4 \pi D(\rho)(T-t) }}\exp\left[
-\frac{(z-x)^{2}}{4D(\rho)(T-t)} \right]
\end{equation}
for $0\le t\le T$. Since $p_{1}(x,t)=0$ at $x\rightarrow -\infty$, we obtain 
\begin{equation}
	p_{1}(x,t)= \frac{1}{\rho}\int_{-\infty}^{x}~dz ~g(0, T\vert z,t)
	=\frac{1}{2\rho}\,\erfc{ \frac{-x}{2\sqrt{D(\rho)\left( T-t
	\right)}} }. \label{eq:p1 solution}
\end{equation}
Taking into account \eqref{eq:dp1 quench}, the solution for $q_{1}(x,t)$ in \eqref{eq:optimal q1} can be written 
as $q_{1}(x,t)=-\partial_{x}\psi(x,t)$ with
\begin{equation}
	\psi(x,t)=\frac{\sigma(\rho)}{\rho}\int_{0}^{t}d\tau\int_{-\infty}^{\infty}dz
	~g(x,t \vert z, \tau)~g(0, T\vert z,\tau).
	\label{eq:psi}
\end{equation}

These are the only two quantities required for simplifying 
$\langle X_{T}^{2}\rangle$ given by Eq.~\eqref{eq:X2}. Recalling that $q_1(x,0)=0$ (and that formally 
$F_{2} \equiv 0$ in the quenched initial state) we obtain
\begin{equation}
	\frac{1}{2!}\left\langle X_{T}^{2} \right \rangle_{\mathcal{Q}}=\frac{1}{\rho}\int_{0}^{\infty} dx 
	~q_{1}(x,T)-\frac{\sigma(\rho)}{2}\int_{0}^{T}dt\int_{-\infty}^{\infty}dx
	\left( \partial_{x}p_{1} \right)^{2}.
	\label{eq:variance quench}
\end{equation}
Using the solution for $q_{1}(x,t)$, the first integral yields
\begin{equation}
	\frac{1}{\rho}\int_{0}^{\infty}dx~q_{1}(x,T)=\frac{\psi(0,T)}{\rho},
\end{equation}
where we have taken into account that $\psi(x,T)$ vanishes at $x\rightarrow \infty$. 
The second integral in \eqref{eq:variance quench} becomes
\begin{equation}
	\frac{\sigma(\rho)}{2}\int_{0}^{T}dt\int_{-\infty}^{\infty}dx\big(\partial_{x}p_{1}\big)^2=\frac{\psi(0,T)}{2\rho}.
\end{equation}
Combining these two results we reduce the expression for the variance to
\begin{equation*}
	\frac{1}{2!} \left\langle X_{T}^{2} \right \rangle_{\mathcal{Q}}=\frac{\psi(0,T)}{2\rho}=\frac{\sigma(\rho)}{2\rho^{2}}\int_{0}^{T}dt\int_{-\infty}^{\infty}dx
	~[(g(0,T\vert x,t )]^2.
\end{equation*}
We use \eqref{g:propagator} and compute the integral.  The final expression for the variance, 
\begin{equation}
	\left\langle X_{T}^{2} \right \rangle_{\mathcal{Q}}=\frac{\sigma(\rho)}{\rho^2}\frac{\sqrt{T}}{
	\sqrt{2\pi D(\rho)}}\,,
	\label{eq:X2 quenched}
\end{equation}
is valid in the quenched setting in the general case of arbitrary $D(\rho)$ and $\sigma(\rho)$. 

\subsection{The annealed initial state \label{eq:variance general}}

Inserting the series expansion \eqref{eq:expansion 1}--\eqref{eq:expansion} into the boundary condition
\eqref{eq:boundary annealed p0}--\eqref{eq:boundary annealed pT} we obtain the boundary conditions in the first order
\begin{equation}
	p_{1}(x,T)=\frac{\Theta(x)}{\rho}\qquad \textrm{and}\qquad
	q_{1}(x,0)=\frac{\sigma(\rho)}{2 D(\rho)}\big( p_{1}(x,0)-p_{1}(x,T) \big]).
	\label{eq:boundary linear annealed}
\end{equation}
The variance \eqref{eq:X2} becomes
\begin{eqnarray}
		\left\langle X_{T}^{2} \right \rangle_{\mathcal{A}}=\frac{1}{\rho}\int_{0}^{\infty}
	\left(q_{1}(x,T)-q_{1}(x,0)\right)dx-\frac{D(\rho)}{\sigma(\rho)}\int_{-\infty}^{\infty}dx
	\left( q_1(x,0) \right)^2\nonumber\\\qquad\qquad\qquad\qquad\qquad-\frac{\sigma(\rho)}{2}\int_{0}^{T}dt\int_{-\infty}^{\infty}dx
	\left( \partial_{x}p_{1}(x,t) \right)^{2},
	\label{eq:variance annealed}
\end{eqnarray}
where we have used \eqref{eq:Y1} and \eqref{eq:F2 annealed}.

The following the analysis is interwoven with the quenched case in a way that allows us to avoid the explicit computation of most integrals. First we note that in both cases the equation for $p_{1}(x,t)$ and the boundary condition on $p_{1}(x,T)$ are identical. Therefore \eqref{eq:p1 solution} remains valid. The equation \eqref{eq:optimal q1} for $q_{1}(x,t)$ is also the same in both cases, but the boundary condition \eqref{eq:boundary linear annealed} is different.  The governing equation \eqref{eq:optimal q1} is linear in $q_{1}(x,t)$, so we can write the solution as a sum 
\begin{equation}
	q_{1}(x,t)=q_{i}(x,t)+q_{h}(x,t),	
\end{equation}	
where $q_{i}(x,t)$ is the solution of the inhomogeneous equation
\begin{equation}
	\partial_{t}q_{i}-D(\rho)\partial_{xx} q_{i}=-
	\sigma(\rho)\partial_{xx}p_{1}
	\qquad\textrm{with}\qquad q_{i}(x,0)=0,
\end{equation}
and $q_{h}(x,t)$ is the solution of the homogeneous equation
\begin{equation}
	\qquad	 \partial_{t}q_{h}-D(\rho)\partial_{xx}
	q_{h}=0\qquad\textrm{with}\qquad
	q_{h}(x,0)=\frac{\sigma(\rho)}{2D(\rho)}\big( p_{1}(x,0)-p_{1}(x,T) \big).
	\label{eq:qh}
\end{equation}

Comparing with the quenched case we notice that $q_{i}(x,t)$ is same as $q_{1}(x,t)$ in the quenched case: $q_i(x,t)=-\partial_x\psi(x,t)$ with $\psi(x,t)$ given by \eqref{eq:psi}. Further, using \eqref{eq:variance quench} and \eqref{eq:variance annealed} we find that the variance for the annealed case is related to the variance in the quenched case via
\begin{equation}
\label{hhh:long}
\frac{1}{2!}\left \langle X_{T}^{2} \right\rangle_{\mathcal{A}} -\frac{1}{2!}\left \langle
	X_{T}^{2} \right\rangle_{\mathcal{Q}}=-\frac{D(\rho)}{\sigma(\rho)}\int_{-\infty}^{\infty}dx
        \big(q_{h}(x,0)\big)^2 +\frac{1}{\rho}\int_{0}^{\infty}dx\,
	\big( q_{h}(x,T)-q_{h}(x,0)\big).
\end{equation}
The difference depends only on the homogeneous solution $q_{h}(x,t)$.

The second term on the right-hand side of Eq.~\eqref{hhh:long} can be simplified using an identity
\begin{equation}
	\frac{1}{\rho}\int_{0}^{\infty}dx\big(q_{h}(x,T)-q_{h}(x,0)\big)
	=\frac{2D(\rho)}{\sigma(\rho)}\int_{-\infty}^{\infty}dx\big(q_{h}(x,0)\big)^2 \, .
\end{equation}
This identity is proved by using \eqref{eq:qh},  then noting that
 $\int_{-\infty}^{\infty}dx\, q_h(x,t)~ p_{1}(x,t)~$ does not depend on time because $ p_{1}$ and $ q_h $ satisfy adjoint equations, and finally that
recalling  $p_{1}(x,T)={\Theta(x)}/{\rho}$:
\begin{eqnarray*}
\frac{2D(\rho)}{\sigma(\rho)}\int_{-\infty}^{\infty}dx\big(q_{h}(x,0)\big)^2  &=&
  \int_{-\infty}^{\infty}dx\ q_{h}(x,0)\big( p_{1}(x,0)-p_{1}(x,T) \big)\\
 &=&  \int_{-\infty}^{\infty}dx \,  p_{1}(x,T) \big(q_{h}(x,T)-q_{h}(x,0)\big).
\end{eqnarray*}
Thus
\begin{eqnarray}
\frac{1}{2!}\left \langle X_{T}^{2} \right\rangle_{\mathcal{A}} -\frac{1}{2!}\left \langle
	X_{T}^{2} \right\rangle_{\mathcal{Q}}
&&=\frac{D(\rho)}{\sigma(\rho)}\int_{-\infty}^{\infty}dx\big(q_{h}(x,0)\big)^2\nonumber\\
&&=\frac{\sigma(\rho)}{4D(\rho)}\int_{-\infty}^{\infty}dx\big(p_{1}(x,T)-p_{1}(x,0)\big)^2,
\end{eqnarray}
where in the last step we have used the boundary condition from Eq.~\eqref{eq:qh}.
The last integral is computed using \eqref{eq:p1 solution} to yield\footnote{The difference $\left \langle X_{T}^{2} \right\rangle_{\mathcal{A}} -\left \langle X_{T}^{2} \right\rangle_{\mathcal{Q}}$ is positive, as shown in Appendix \ref{app:inequality}.}
\begin{equation}
	\frac{1}{2!}\left \langle X_{T}^{2} \right\rangle_{\mathcal{A}} -\frac{1}{2!}\left \langle
	X_{T}^{2} \right\rangle_{\mathcal{Q}}
	=\frac{\sigma(\rho)}{2\rho^{2}}\left(
	\frac{\sqrt{2}-1}{\sqrt{2\pi}} \right)\sqrt{\frac{T}{D(\rho)}}.
	\label{eq:difference}
\end{equation}
Combining this with \eqref{eq:X2 quenched} we establish a very simple general relation
\begin{equation}
	\left \langle X_{T}^{2} \right\rangle_{\mathcal{A}}=\sqrt{2}~\left \langle
	X_{T}^{2} \right\rangle_{\mathcal{Q}}
\end{equation}
between the two initial states. This relation and  Eq.~\eqref{eq:X2 quenched} are applicable for any single-file diffusion, i.e., for arbitrary transport coefficients.\footnote{Our derivations are based on the MFT which assumes local equilibrium. This is expected to be correct for systems with short-range inter-particle interactions.} To verify the range of applicability of Eq.~\eqref{eq:X2 quenched} let us compare with earlier work. In the simplest case of Brownian particles we recover the well-known expression for the variance (for more about Brownian particles see Refs.~\cite{Harris1965,Rodenbeck1998,Lizana2008,Leibovich2013}).  For the SEP we also reproduce the well-known result \cite{Arratia1983}. Let us also compare with the variance in single-file colloidal systems studied in Ref.~\cite{Kollmann2003}. For comparison, we need the relation
\begin{equation}
\sigma(\rho)=2S(0,0)D(\rho),
\end{equation}
where $S(0,0)$ is the structure factor \cite{Spohn1991}. With this the variance for the annealed case becomes
\begin{equation}
\left \langle X_{T}^{2} \right\rangle_{\mathcal{A}}=\frac{2S(0,0)}{\rho^2}\sqrt{\frac{D(\rho)}{\pi}}~\sqrt{T},
\end{equation}
which is identical to the one presented in \cite{Kollmann2003}. For experimental measurements, it is convenient to express the variance in terms of the  isothermal compressibility $\kappa$. Since $S(0,0)=\beta^{-1}\kappa ~\rho^2$, \begin{equation}
\left \langle X_{T}^{2} \right\rangle_{\mathcal{A}}=2~\beta^{-1}~\kappa ~\sqrt{\frac{D(\rho)}{\pi}}~\sqrt{T},
\end{equation}
where $\beta$ is the inverse temperature.

\section{Fourth cumulant of a tagged particle in the SEP   \label{sec:fourth}}

The symmetric exclusion process (SEP) is a diffusive lattice gas. Each site is occupied by at most one particle and each particle attempts to hop to neighboring empty sites with unit rate. For the SEP,  
the diffusion coefficient is constant, $D(\rho)=1$, while the mobility has the symmetric form $\sigma(\rho)=2\rho (1-\rho)$ reflecting the $\rho\leftrightarrow 1-\rho$ mirror symmetry of the SEP. The  formula for $\sigma(\rho)$ can be derived using the fluctuation-dissipation relation (see  \aref{sec:sigma derivation}). 

We employ a perturbative approach and compute the fourth cumulant of  a tagged particle position. It proves useful to consider a one-parameter class of models with $D(\rho)=1$ and $\sigma(\rho)=2\rho(1-\alpha\rho)$. The 
parameter $\alpha$ varies in the range $0\le \alpha \le 1$, so the mobility remains positive. The case of $\alpha=0$ describes the single-file system of Brownian particles, whereas $\alpha=1$ corresponds to the SEP.

Mathematically, we want to solve  equations \eqref{eq:opt p}--\eqref{eq:opt q}. The corresponding boundary conditions depend on the solution itself. As discussed in \sref{sec:point particle}, a way to solve the equations is by treating $B=\lambda/q(Y,T)$ as a parameter. This determines the optimal fields in terms of $B$ which we denote by $q_B(x,t)$ and $p_B(x,t)$. The solution does not explicitly involve the fugacity parameter $\lambda$ which only appears through the dependence of $B$ on $\lambda$. A straightforward implementation of the definition of $B$ does not lead to a unique value, as the function $q(x,T)$ is singular at $(x,t)\equiv(Y,T)$. The value corresponding to the least action path is obtained by an optimization condition $\frac{d\mu}{dB}=0$.
This expresses $B$ in terms of $\lambda$ and leads to a parametric formula of the cumulant generating function
\begin{equation}
\mu=\lambda ~Y(B)-R(B) \qquad \textrm{with} \qquad \lambda~\dfrac{d Y(B)}{dB}=\dfrac{dR(B)}{dB}.
\label{eq:muB parameter}
\end{equation}
The function $R(B)$ is defined in terms of the optimal fields $q_B(x,t)$ and $p_B(x,t)$, Eq.~\eqref{eq:mu formal Quench} in the quenched case and Eq.~\eqref{eq:mu formal annealed} in the annealed case. A detailed implementation of this procedure has been presented in \sref{sec:point particle} for the system of Brownian particles where an exact solution was possible as the corresponding $\sigma(\rho)$ is linear.

To proceed with the analysis for the quadratic $\sigma(\rho)$, it is instructive to recall  the symmetry properties of the optimal fields \eqref{eq:symmetry q}-\eqref{eq:symmetry mu}: 
\begin{equation*}
q_{-B}(-x,t)= q_B(x,t) \qquad \textrm{and} \qquad 
p_{-B}(-x,t)= p_B(x,t)-B,
\end{equation*}
whereas $Y(B)$ and $\lambda(B)$ are odd functions of $B$. Combining all together shows that $R(B)$ is an even function of $B$. This is consistent with the fact that the dynamics is unbiased.

It proves convenient to use $B$ as a primary expansion parameter; at the end one can re-expand the results in terms of  $\lambda$ and compute the cumulants. The aforementioned symmetry properties allow us to seek $Y(B)$, $R(B)$ and $\lambda(B)$ as the following expansions
\begin{equation}
\label{eq:Y series B}
\begin{split}
Y(B)&= Y_1 B+Y_3 B^3+\mathcal{O}(B^5)\\
R(B)&= R_2 B^2+R_4 B^4+\mathcal{O}(B^6)\\
\lambda(B)&= \lambda_1 B+\lambda_3 B^3+\mathcal{O}(B^5).
\end{split}
\end{equation}
The last formula is also equivalent to
\begin{equation}
B=\left(\dfrac{1}{\lambda_1}\right)\lambda-\left(\dfrac{\lambda_3}{\lambda_1^4}\right)\lambda^3+\mathcal{O}(\lambda^5).
\end{equation}

Substituting the series expansions in the first  equation of   \eqref{eq:muB parameter} leads to a formula for the cumulant generating function in powers of $B$, 
\begin{equation*}
\mu=\big( \lambda_1 Y_1-R_2 \big)B^2+\big(\lambda_1Y_3+\lambda_3 Y_1-R_4 \big)B^4+\cdots \, .
\end{equation*}
The second equation of  \eqref{eq:muB parameter}, rewritten as
 $ \lambda = \dfrac{dR}{dB} \left(\dfrac{dY}{dB}\right)^{-1} $ leads to 
\begin{equation}
\lambda_1=\dfrac{2R_2}{Y_1} \qquad \textrm{and} \qquad \lambda_3=\dfrac{4R_4-3\lambda_1Y_3}{Y_1}.
\end{equation}
To determine the cumulants we use the above in the series expansion  \eqref{eq:mu expansion cumulants} and obtain
\begin{align}
\left\langle X_{T}^{2} \right \rangle_{c}&= \dfrac{Y_1}{\lambda_1},\label{eq:formal Y2}\\
\dfrac{1}{4!}\left\langle X_{T}^{4} \right \rangle_{c}&= \dfrac{1}{\lambda_1^4}\big(-  R_4+\lambda_1 Y_3\big).
\label{eq:formal Y4}
\end{align}
These formal expressions hold in both annealed and quenched cases. 

The boundary condition for the optimal fields derived in \eqref{eq:boundary annealed p0} and \eqref{eq:boundary annealed pT} can be rewritten as
\begin{align}
p_B(x,0)&= B~\Theta(x)+\int_{\rho}^{q_{B}(x,0)}dr ~\dfrac{2D(r)}{\sigma(r)}
 =  B~\Theta(x)+ \ln \frac{q_{B}(x,0) ( 1 - \alpha \rho)}{\rho( 1 - \alpha q_{B}(x,0)) }  
  \label{eq:boundary p0 in terms of B}\\
p_B(x,T)&= B~\Theta(x-Y),
\label{eq:boundary in terms of B}
\end{align}
where we used the definition of $B$ in \eqref{eq:B def}.
To proceed, we write an expansion of the optimal fields in powers of $B$ as
\begin{align}
q_B&= \rho+q_1 B+q_2 B^2+q_3 B^3+\cdots, \label{eq:series q in B}\\
p_B&= p_1 B+p_2B^2+p_3 B^3+\cdots.  \label{eq:series p in B}
\end{align}
Note that the $q_k(x,t)$ and $p_k(x,t)$ in the above formulas are different from those in \eqref{eq:expansion 1}--\eqref{eq:expansion} which were obtained using expansions in powers of $\lambda$, and not $B$.

A straightforward computation of the solution to different orders is tedious, it involves difficult integrals. We circumvent this by drawing comparison with the $\alpha=0$ case where an exact solution is available. We first illustrate this trick by computing the solution in the linear order in $B$. This will give us the second cumulant which we can compare with already known results (which were derived in the previous section in the general setup). Then we shall compute the forth cumulant. 

\subsubsection*{The second cumulant}
The governing equations in the first order in $B$ are
\begin{align*}
\big(\partial_t+\partial_{xx}\big)p_1&= 0,\\
\big(\partial_t-\partial_{xx}\big)q_1&= -2\rho\big( 1-\alpha \rho \big) \partial_{xx} p_1.
\end{align*}
To determine the corresponding boundary conditions we use a formal expansion of the step function,
\begin{equation}
\Theta(x-Y)=\Theta(x)-\delta(x)Y+\dfrac{1}{2}\delta^{\prime}(x)Y^2+\cdots.
\end{equation}
The boundary conditions are
\begin{equation}
p_1(x,T)=\Theta(x) \qquad \textrm{and} \qquad q_1(x,0)=\rho~\big(1-\alpha \rho \big)~\big(p_1(x,0)-p_1(x,T)\big).
\end{equation}
The solutions to the governing equations are almost the same as in the Brownian case:
\begin{equation}
p_1(x,t)=\widehat{p}_1(x,t)\qquad \textrm{and} \qquad q_1(x,t)=\big(1-\alpha \rho \big)~ \widehat{q}_1(x,t),
\label{eq:p1 q1 relation}
\end{equation}
where the hat denotes the corresponding solutions for $\alpha=0$ case, \textit{i.e.}, for Brownian particles. In the rest of this paper we shall follow the same notation.

To derive the second cumulant we need to determine $Y_1, \lambda_1$ and $R_2$. Combining the series expansion for $Y$ with the definition \eqref{eq:Yconverge} results in
\begin{equation}
Y_1=\dfrac{1}{\rho}\int_{0}^{\infty}dx \big( q_1(x,T)-q_1(x,0)\big).
\end{equation}
Thanks to \eqref{eq:p1 q1 relation} we have 
\begin{equation}
Y_1=\big( 1-\alpha \rho \big) ~\widehat{Y}_1.
\label{eq:Y1 relation}
\end{equation}

To compute $R_2$, we start with the general formula \eqref{eq:mu formal annealed} which in the present case becomes
\begin{equation}
R=\int_{-\infty}^{\infty}dx\int_{\rho}^{q(x,0)}dr
~\frac{1}{r(1-\alpha r)} \big( q(x,0)-r \big)+\int_{0}^{T}dt\int_{-\infty}^{\infty} dx~q\big(1-\alpha q\big)\big(\partial_x p\big)^2.
\label{eq:R annealed}
\end{equation}
At second order we get
\begin{equation}
R_2=\dfrac{1}{2 \rho(1-\alpha \rho)}\int_{-\infty}^{\infty}dx~\big(q_1(x,0)\big)^2+   \rho\big(1-\alpha \rho\big)\int_{0}^{T}dt\int_{-\infty}^{\infty} dx~\big(\partial_x p_1\big)^2
\end{equation}
which in conjunction with \eqref{eq:p1 q1 relation} yields 
\begin{equation}
R_2=\big(1-\alpha\rho\big)\widehat{R}_2.
\label{eq:R2 relation}
\end{equation}
The above relations \eqref{eq:Y1 relation} and \eqref{eq:R2 relation} also establish
\begin{equation}
\label{LL}
\lambda_1=\widehat{\lambda}_1.
\end{equation}
Using \eqref{eq:Y1 relation} and \eqref{LL} we transform \eqref{eq:formal Y2} into
\begin{equation}
 \left\langle X_T^2 \right\rangle_{\mathcal{A}}=(1-\alpha \rho)\left\langle \widehat{X}_T^2 \right\rangle_{\mathcal{A}}.
\end{equation}
The variance $\langle \widehat{X}_T^2\rangle_{\mathcal{A}}$ for the Brownian case is derived in \eqref{eq:X2 annealed point} which then leads to 
\begin{equation}
\left\langle X_T^2 \right\rangle_{\mathcal{A}}=2\,\dfrac{1-\alpha \rho}{\rho}\,\dfrac{\sqrt{T}}{\sqrt{\pi}}.
\end{equation}
This expression agrees with the general formula for the variance derived earlier in \sref{eq:variance general}, validating the approach we used here.

The variance can also be determined by directly solving for $q_1$ and $p_1$. An extension of this approach to the fourth cumulant requires explicit solution of the optimal fields up to the third order which is a very tedious task. The alternative approach based on the mapping to $\alpha=0$ case, as demonstrated above for the variance, considerably simplifies computations, so we adopt this approach in the following derivation of the fourth cumulant.
Before proceeding, we write explicit formulas $\lambda_1$ and $Y_1$. Using \eqref{eq:lambda annealed point final} which describes the case of $\alpha=0$, we compute $\widehat{\lambda}_1=\rho$ leading [due to \eqref{LL}] to 
\begin{equation}
\lambda_1=\rho.
\end{equation}
Similarly we derive
\begin{equation}
Y_1=(1-\alpha\rho)~\dfrac{2}{\sqrt{\pi}}~\sqrt{T}.
\label{eq:final Y1}
\end{equation}

\subsubsection*{The fourth cumulant}

In the second order in $B$ we have
\begin{align*}
\big(\partial_t+\partial_{xx}\big)p_2&=-\big(1-2\alpha\rho\big)
\big(\partial_xp_1\big)^2,\\
\big(\partial_t-\partial_{xx}\big)q_2&=-2\rho\big(1-\alpha \rho \big)\partial_{xx}p_2-2\big(1-2\alpha \rho\big)\partial_x\big(q_1\partial_xp_1\big),
\end{align*}
whereas the corresponding equations for the third order are
\begin{align*}
\big(\partial_t+\partial_{xx}\big)p_3&=-2\big(1-2\alpha\rho\big)
\big(\partial_xp_1\big)\big(\partial_xp_2\big)+2\alpha q_1\big(\partial_xp_1\big)^2,\\
\big(\partial_t-\partial_{xx}\big)q_3&=-\partial_x\Bigg[2\rho\big(1-\alpha \rho \big)\partial_{x}p_3+2\big(1-2\alpha \rho\big)q_1\partial_xp_2+2\big(\big(1-2\alpha\rho\big)q_2-\alpha\big(q_1
\big)^2\big)\partial_xp_1\Bigg].
\end{align*}
These equation are derived from  the Hamilton equations \eqref{eq:opt p}--\eqref{eq:opt q}. 
The corresponding boundary conditions follow from \eqref{eq:boundary p0 in terms of B}--\eqref{eq:boundary in terms of B}: 
\begin{align*}
p_2(x,T)&= -Y_1~\delta(x),\\
q_2(x,0)&= \rho(1-\alpha\rho)p_2(x,0) + \dfrac{\rho(1-\alpha\rho)(1-2\alpha\rho)}{2}\left[p_1(x,T)-p_1(x,0)\right]^2,\\
p_3(x,T)&= \dfrac{1}{2} Y_1^2 ~\delta^{\prime}(x),\\
q_3(x,0)&= \rho (1-\alpha\rho)p_3(x,0)+\dfrac{1-2\alpha\rho}{\rho(1-\alpha \rho)}\,q_1(x,0) q_2(x,0)-\frac{1-3\alpha \rho(1-\alpha \rho)}{3[\rho(1-\alpha \rho)]^2}\, [q_1(x,0)]^3.
\end{align*}

From the above formulas, one can verify that the solutions at the second order are related to the corresponding solutions for $\alpha=0$ case by a simple transformation:
\begin{align}
p_2(x,t)&= \big(1-2\alpha\rho\big)~\widehat{p}_2(x,t)\Big\vert_{\widehat{Y}_1\rightarrow\tfrac{Y_1}{(1-2\alpha \rho)}} \label{eq:p2 relation},\\
q_2(x,t)&=\big(1-\alpha \rho\big) \big(1-2\alpha\rho\big)~\widehat{q}_2(x,t)\Big\vert_{\widehat{Y}_1\rightarrow\tfrac{Y_1}{(1-2\alpha \rho)}}.
\end{align}
On the right-hand side we write $\widehat{Y}_1\rightarrow\tfrac{Y_1}{(1-2\alpha \rho)}$ implying that $\widehat{Y}_1$ should be replaced by $\tfrac{Y_1}{(1-2\alpha \rho)}$ in the corresponding solution for $\alpha=0$ case obtained by treating $\widehat{Y}_1$ as parameter.

A similar relation can be derived for the third order terms:
\begin{align}
p_3(x,t)&= \big(1-2\alpha\rho\big)^2~\widehat{p}_3(x,t)\Big\vert_{\widehat{Y}_1\rightarrow\tfrac{Y_1}{(1-2\alpha \rho)}}+\alpha\big(1-\alpha \rho\big) u(x,t),\\
q_3(x,t)&=\big(1-\alpha \rho\big) \big(1-2\alpha\rho\big)^2~\widehat{q}_3(x,t)\Big\vert_{\widehat{Y}_1\rightarrow\tfrac{Y_1}{(1-2\alpha \rho)}}+\alpha\big(1-\alpha\rho\big)^2 h(x,t).
\end{align}
Here $u(x,t)$ is the solution of 
\begin{equation}
\label{eq:u eq}
\big(\partial_t+\partial_{xx}\big)u = 2~\widehat{q}_1~\big(\partial_x\widehat{p}_1\big) 
\end{equation}
subject to $u(x,T)=0$, while $h(x,t)$ is the solution of 
\begin{align}
\label{eq:h eq}
\big(\partial_t-\partial_{xx}\big)h =-2~\rho~\partial_{xx}u
+2~\partial_x\left[\widehat{q}_1^2~\partial_x\widehat{p}_1\right]
\end{align}
subject to $h(x,0)=\rho ~u(x,0)-\dfrac{1}{3\rho}\left[\widehat{q}_1(x,0) \right]^3$. Note that both $u(x,t)$ and $h(x,t)$ do not depend on $\alpha$. These functions also appear in the analysis of the statistics of time integrated current in the symmetric exclusion process on an infinite line (see \aref{sec:current}).

In order to compute the fourth cumulant we must find $R_4$ and $Y_3$.  Combining the series expansion  \eqref{eq:series q in B}--\eqref{eq:series p in B} and \eqref{eq:R annealed} we get
\begin{equation*}
R_4=\int_{-\infty}^{\infty}dx\,U_4(x)+\int_{0}^{T}dt\int_{-\infty}^{\infty}dx\,V_4(x,t)
\end{equation*}
with
\begin{align*}
U_4 =~&\dfrac{[q_2(x,0)]^2+ 2q_1(x,0)q_3(x,0)}{2\rho(1-\alpha \rho)}-\dfrac{(1-2\alpha\rho)}{2\rho^2(1-\alpha\rho)^2}\big(q_1(x,0)\big)^2q_2(x,0)\\
&+\dfrac{(1-2\alpha\rho)^2}{12\rho^3(1-\alpha\rho)^3}\big(q_1(x,0)\big)^4
+\dfrac{\alpha}{12\rho^2(1-\alpha\rho)^2}\big(q_1(x,0)\big)^4\\
V_4 =~&
\rho(1-\alpha\rho)\left[\big(\partial_x p_2\big)^2 + 2\big(\partial_xp_1\big)\big(\partial_xp_3\big)\right] 
-\alpha q_1^2\big(\partial_xp_1\big)^2\\
&+(1-2\alpha\rho)\left[q_2\big(\partial_xp_1\big)^2+2q_1\big(\partial_xp_1\big)
\big(\partial_xp_2\big)\right] .
\end{align*}

We now again express $R_4$ through the corresponding solution for $\alpha=0$ case:
\begin{equation}
R_4=\big(1-\alpha\rho\big)\big(1-2\alpha \rho \big)^2 \Bigg[\widehat{R}_4\Big\vert_{\widehat{Y}_1\rightarrow\frac{Y_1}{1-2\alpha\rho}}\Bigg]+\rho^2\alpha\big(1-\alpha\rho\big)^2~\mathcal{I}.
\label{R4avec hat}
\end{equation}
The last term is given by 
\begin{align*}
\mathcal{I}&=\int_{-\infty}^{\infty}dx\Bigg[\bigg(\dfrac{\widehat{q}_1(x,0)}{\rho}\bigg)\bigg(\dfrac{h(x,0)}{\rho^2}\bigg)+\dfrac{1}{12}\bigg(\dfrac{\widehat{q}_1(x,0)}{\rho}\bigg)^4\Bigg]\\
& + \int_{0}^{T}dt\int_{-\infty}^{\infty}dx\Bigg[2\big(
\partial_x\widehat{p}_1\big)\left(\dfrac{\partial_x u}{\rho}\right)-\left(\dfrac{\widehat{q}_1}{\rho}\right)^2\big(\partial_x\widehat{p}_1\big)^2\Bigg].
\end{align*}
This only involves the solution for the $\alpha=0$ case.

The second quantity required to compute the fourth cumulant is  $Y_3$. Combining the series expansion \eqref{eq:Y series B} and \eqref{eq:Yconverge}, and using the relation of the optimal fields to their counterparts for $\alpha=0$,
we can express $Y_3$  in terms of the hat variables:
\begin{equation}
Y_3=\big(1-\alpha \rho\big)\big(1-2\alpha\rho\big)^2~\widehat{Y}_3\Big\vert_{\widehat{Y}_1\rightarrow\frac{Y_1}{1-2\alpha\rho}}+\alpha\rho \big(1-\alpha \rho \big)^2\int_{0}^{\infty}dx\Bigg[\dfrac{{h}(x,T)}{\rho^2}-\dfrac{{h}(x,0)}{\rho^2}\Bigg].
\end{equation}
Combining this equation with \eqref{R4avec hat}, we find that $Y_3-R_4/\rho$ which appears in the fourth cumulant  \eqref{eq:formal Y4} can be written as
\begin{eqnarray}
-\dfrac{R_4}{\rho}+Y_3 &=&(1-\alpha\rho)(1-2\alpha\rho)^2~\Bigg[-\dfrac{\widehat{R}_4}{\rho}+\widehat{Y}_3\Bigg]\Bigg\vert_{\widehat{Y}_1\rightarrow\frac{Y_1}{1-2\alpha\rho}}\nonumber\\
&+&\alpha\rho\big(1-\alpha\rho\big)^2
\Bigg[- \mathcal{I}+\int_{0}^{\infty}dx\Bigg(\dfrac{{h}(x,T)}{\rho^2}-\dfrac{{h}(x,0)}{\rho^2}\Bigg)\Bigg].
\label{eq:mapping one}
\end{eqnarray}
The term inside the square brackets in the bottom line of \eqref{eq:mapping one} simplifies to
\begin{equation}
- \mathcal{I}+\int_{0}^{\infty}dx \Bigg(\dfrac{{h}(x,T)}{\rho^2}-\dfrac{{h}(x,0)}{\rho^2}\Bigg)=\dfrac{1}{4!}\Bigg(\dfrac{8-6\sqrt{2}}{\sqrt{\pi}}\Bigg)\sqrt{T}.
\label{eq:integral one}
\end{equation}
This identity was verified numerically using Mathematica. We also obtained an analytical proof by comparing to the  analysis of the current in the SEP with uniform initial profile (see \aref{sec:current}).

On the other hand, from the exact solution \eqref{eq:mu Ann final} for the $\alpha=0$ case we get (by substituting $Y(B)$ from \eqref{eq:Y series B} in  \eqref{eq:mu Ann final} and expanding in powers of B)
\begin{equation}
-\dfrac{\widehat{R}_4}{\rho}+\widehat{Y}_3=\dfrac{\sqrt{T}}{12\sqrt{\pi}}-\dfrac{\widehat{Y}_1}{6}+\dfrac{\widehat{Y}_1^2}{4\sqrt{T\pi}}\,.
\end{equation}
This allows us to simplify the term inside the square brackets in the top line of \eqref{eq:mapping one}
\begin{equation}
\Bigg[-\dfrac{\widehat{R}_4}{\rho}+\widehat{Y}_3\Bigg]\Bigg\vert_{\widehat{Y}_1\rightarrow\frac{Y_1}{1-2\alpha\rho}}=\dfrac{\sqrt{T}}{12\sqrt{\pi}}-\dfrac{Y_1}{6(1-2\alpha\rho)}+\dfrac{Y_1^2}{4\sqrt{T \pi}(1-2\alpha\rho)^2}.
\end{equation}
Substituting $Y_1$ from \eqref{eq:final Y1}, we obtain 
\begin{equation}
\frac{1}{\sqrt{T}}\Bigg[-\dfrac{\widehat{R}_4}{\rho}+\widehat{Y}_3\Bigg]\Bigg\vert_{\widehat{Y}_1\rightarrow\frac{Y_1}{1-2\alpha\rho}}=\dfrac{1}{12\sqrt{\pi}}-\dfrac{(1-\alpha\rho)}{(1-2\alpha\rho)}~\dfrac{1}{3\sqrt{\pi}}+\dfrac{(1-\alpha\rho)^2}{(1-2\alpha\rho)^2}\dfrac{1}{\pi^{3/2}}.
\end{equation}
After all this work Eq.~\eqref{eq:mapping one} becomes
\begin{eqnarray}
\label{long:4}
\frac{1}{\sqrt{T}}\Bigg[-\dfrac{R_4}{\rho}+Y_3\Bigg]&=& (1-\alpha\rho)(1-2\alpha\rho)^2~\dfrac{1}{12\sqrt{\pi}}-(1-\alpha\rho)^2(1-2\alpha\rho)~\dfrac{1}{3\sqrt{\pi}} \nonumber\\
&+&(1-\alpha\rho)^3~\dfrac{1}{\pi^{3/2}}+\dfrac{\alpha\rho(1-\alpha\rho)^2}{4!}\Bigg(\dfrac{8-6\sqrt{2}}{\sqrt{\pi}}\Bigg).
\end{eqnarray}
Plugging \eqref{long:4} and $\lambda_1=\rho$ into \eqref{eq:formal Y4} we obtain
\begin{equation}
\label{4:cumulant}
\left\langle X_T^4 \right\rangle_{c}=\dfrac{(1-\alpha\rho)}{\rho^3}~\Bigg[1-\bigg(4-\big(8-3\sqrt{2}\big)\alpha\rho\bigg)
\big(1-\alpha \rho\big)+\dfrac{12}{\pi}\big(1-\alpha\rho\big)^2\Bigg]~\dfrac{2\sqrt{T}}{\sqrt{\pi}}.
\end{equation}
Specifying \eqref{4:cumulant} to $\alpha=0$ we get back the result \eqref{eq:X4 annealed point}, whereas setting $\alpha=1$ we arrive at the fourth cumulant for the SEP
\begin{equation}
\label{4:SEP}
\left\langle X_T^4 \right\rangle_{c}=\dfrac{1-\rho}{\rho^3}~\Bigg[1-\bigg(4-\big(8-3\sqrt{2}\big)\rho\bigg)
\big(1-\rho\big)+\dfrac{12}{\pi}\,\big(1-\rho\big)^2\Bigg]~\dfrac{2\sqrt{T}}{\sqrt{\pi}}.
\end{equation}
This result, announced in Ref.~\cite{KMS2014}, is valid in
the annealed setting.\footnote{We also  derived  (\ref{4:SEP})  by  a
 straightforward perturbative calculation, without introducing the 
interpolating parameter $\alpha$.}  When
 $\rho\rightarrow 1$, this expression matches the result derived in Ref.~\cite{Illien2013}.

\paragraph{Remark:} The fourth cumulant in the quenched setting can be calculated along similar
lines to yield
\begin{equation*}
\left \langle X_T^4 \right \rangle_{c}^{\mathcal{Q}}=\dfrac{1-\alpha\rho}{\rho^3}~\Bigg[2(1-2\alpha \rho)^2\bigg(\frac{9}{\pi}\arctan\left(\frac{1}{2\sqrt{2}}\right)-1\bigg)+\alpha \rho\big(1-\alpha\rho\big)\big(4-3\sqrt{2}\big)\Bigg]\sqrt{\frac{2}{\pi}}~\sqrt{T}.
\end{equation*}
For Brownian particles, $\alpha=0$, we recover Eq.~\eqref{cumulant4:Bquenched}. Setting  $\alpha=1$, we obtain the fourth cumulant for the SEP in the quenched case:
\begin{equation*}
\left \langle X_T^4 \right \rangle_{c}^{\mathcal{Q}}=
(1-\rho)\left[2\,\frac{(1-2\rho)^2}{\rho^3}\bigg(\frac{9}{\pi}\arctan\left(\frac{1}{2\sqrt{2}}\right)-1\bigg)
+\frac{1-\rho}{\rho^2}\,\big(4-3\sqrt{2}\big)\right]\sqrt{\frac{2}{\pi}}~\sqrt{T}.
\end{equation*}

\section{A microscopic derivation for the Brownian point particles. \label{sec:microscopic}}

The microscopic  problem was first studied  by Harris \cite{Harris1965}, who derived an exact formula for the variance of the tagged particle. The analysis used the fact that trajectories of the particles are related to the trajectories of non-interacting particles with an exchange of particle index to keep the ordering same
(see  \cite{Hegde2014} for a recent reference).

There is an equivalent description of the problem in terms of the phase space trajectories 
\cite{DeepakKumar2008,Lizana2008,Rodenbeck1998}. Consider $2n+1$ point particles diffusing on a one-dimensional line. The only interaction between particles is the hard-core repulsion which preserves the order of the particles. The particles are indexed by $\left\{-n,-n+1,\cdots,n \right\}$. The central particle is set to be the tagged particle. Let  $\mathbf{Y}\equiv\left\{
y_{-n},\cdots,y_{n} \right\}$, with $-L\le y_{-n}<\cdots<y_{n}\le L$, be the positions at time $t=0$. The positions at time $t$ are denoted by 
$\mathbf{X}\equiv\left\{ x_{-n},\cdots,x_{n} \right\}$.

\begin{figure}
\begin{center}
\includegraphics[scale=0.75]{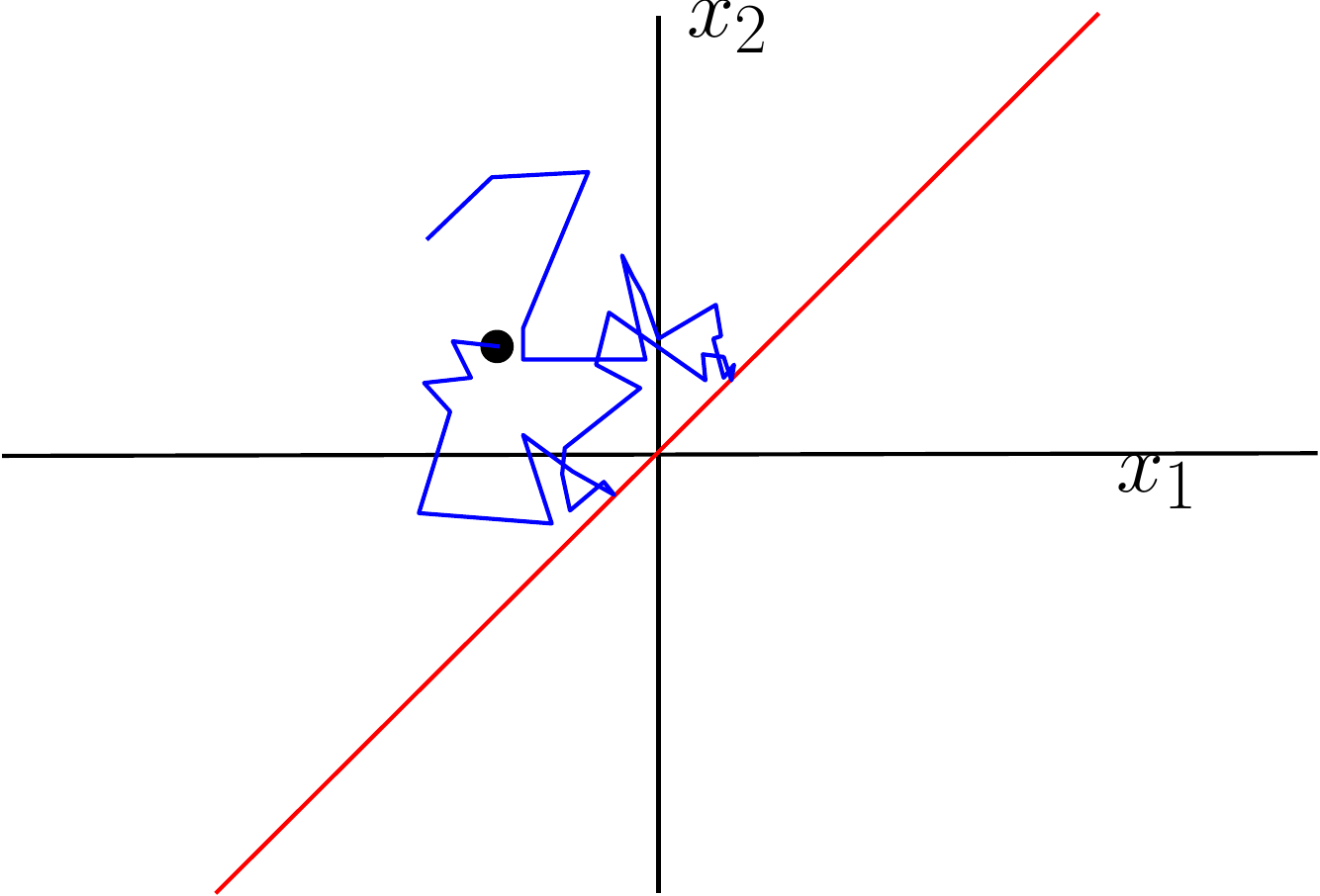}
\end{center}
\caption{A schematic representation of a sample trajectory of two one-dimensional Brownian point particles on the coordinate plane $(x_1,x_2)$. Due to the hard-core repulsion between particles the motion is confined in the domain $x_2>x_1$, whose boundary is denoted by the diagonal line.  \label{fig:weyl chamber}}
\end{figure}

The evolution of the particles can be described by diffusion in a $2n+1$ dimensional space confined to the chamber $x_{-n}<x_{-n+1}<\cdots<x_{n}$ (see \fref{fig:weyl chamber} for a schematic). The probability $P_{t}(\X\vert \Y)$ of the particle position follows a diffusion equation
\begin{equation}
	\partial_{t}P_{t}(\X\vert
	\Y)= \sum_{i=-n}^{n}\partial_{x_{i}}^{2}P_{t}(\X\vert \Y)
	\label{eq:diff eq}
\end{equation}
where the diffusion coefficient is again set to unity. The single-file constraint is implemented by a reflecting boundary condition along the boundary of the chamber:
\begin{equation}
	\partial_{x_{i}}P_{t}(\X\vert \Y)=\partial_{x_{i+1}}P_{t}(\X\vert
	\Y)\qquad\textrm{at}\quad x_{i}=x_{i+1} \textrm{ for all }-n\le i <n.
	\label{eq:sf boundary}
\end{equation}

The solution to Eqs.~\eqref{eq:diff eq}--\eqref{eq:sf boundary} can be written as
\begin{equation}
	P_{t}(\X\vert
	\Y)=\int_{-\infty}^{\infty}\frac{dk_{-n}}{2\pi}\cdots\int_{-\infty}^{\infty}\frac{dk_{n}}{2\pi}{\displaystyle\sum_{\sigma}}\prod_{j=-n}^{n}e^{- 
	t k_{j}^{2}+i k_{j}\left( x_{\sigma(j)}-y_{j} \right) },
	\label{eq:ansatz 1}
\end{equation}
where $\sigma$ is the permutation operator acting on the indexes of the particles. The above solution is applicable only within the chamber $ x_{-n}<x_{-n+1}<\cdots<x_{n}$. Equation \eqref{eq:ansatz 1} is one of the simplest examples of the Bethe ansatz.

The Gaussian integrals in \eqref{eq:ansatz 1} can be evaluated yielding
\begin{equation}
	P_{t}(\X\vert
	\Y)=\sum_{\sigma}\prod_{j=-n}^{n}g_{t}(x_{j}\vert y_{\sigma(j)}), \qquad 
	g_{t}(x\vert y)\equiv \frac{1}{\sqrt{4\pi   t}}e^{-\frac{\left(
	x-y\right)^{2}}{4   t}}\,.
	\label{eq:ansatz 2}
\end{equation}

\subsection{Probability of the tagged particle position}
The central particle ($j=0$) is chosen to be the tagged particle. Without loss of generality, we assume that the tagged particle starts at the origin, $y_{0}=0$. The probability of finding the tagged particle at $x_{0}=x$ at time $t$ is 
\begin{equation*}
	\text{Prob}_{t}(x\vert \Y) =
	\int_{-\infty}^{x_{-n+1}}dx_{-n}\int_{-\infty}^{x_{-n+2}}dx_{-n+1}\ldots\int_{-\infty}^{x}dx_{-1} \int_{x}^{\infty}dx_{1}\int_{x_{1}}^{\infty}dx_{2}\ldots\int_{x_{n-1}}^{\infty}dx_{n}
~P_{t}(\X\vert\Y).
\end{equation*}
This function has been studied in great details by R\"{o}denbeck \textit{et al.} \cite{Rodenbeck1998} for the annealed initial condition. Their results allow one to extract the large deviation function. Here we present a short alternative derivation, and then we determine the large deviation in the quenched case.

Let
\begin{equation}
	G\left(x, y_1,\cdots,y_{n}\right)=\int_{x}^{\infty}dz_{1}\int_{z_{1}}^{\infty}dz_{2}\cdots
	\int_{z_{n-1}}^{\infty}dz_{n}\prod_{j=1}^{n}g_{t}(z_{j}\vert y_{j}).
	\label{eq:G def}
\end{equation}
In terms of this function 
\begin{equation*}
	 	\text{Prob}_{t}(x\vert \Y)=\sum_{\sigma}g_{t}(x\vert
	y_{\sigma(0)})~ G\left(
	-x,-y_{\sigma(-1)},-y_{\sigma(-2)},\cdots,-y_{\sigma(-n)}
	\right)G\left(x,y_{\sigma(1)},y_{\sigma(2)},\cdots,y_{\sigma(n)}\right).
\end{equation*}
Let us rewrite the above formula as
\begin{eqnarray}
	\text{Prob}_{t}(x\vert \Y)=\sum_{k=-n}^{n}g_{t}(x\vert
		y_{k}) A_{k}\left[ x,\Y_{k} \right],
		\label{eq:prob exact 0}
\end{eqnarray}
where we grouped the terms with same values of $\sigma(0)=k$. Further, $\Y_{k}$ denotes the subset of $\Y$ excluding the $k$-th element $y_{k}$. The
amplitude is
\begin{eqnarray}
		A_{k}\left[ x,\Y_{k} \right]=\sum_{\sigma_{k}}G\left(
	-x,-y_{\sigma_{k}(-1)},-y_{\sigma_{k}(-2)},\cdots,-y_{\sigma_{k}(-n)}
	\right) G\left(x,y_{\sigma_{k}(1)},y_{\sigma_{k}(2)},\cdots,y_{\sigma_{k}(n)}\right),
	\label{eq:amplitude}
\end{eqnarray}
where the permutation operator $\sigma_{k}$ denotes permutations acting on the
subset $\Y_{k}$.

Equation \eqref{eq:amplitude} can be simplified using an identity \eqref{eq:id 0}, proved in the Appendix \ref{app:identity 1 G}, resulting in
\begin{eqnarray}
	A_{k}\left[ x,\Y_{k}
	\right]=\sum_{\epsilon_{-n}}\cdots \sum_{\epsilon_{k-1}}\sum_{\epsilon_{k+1}} \cdots \sum_{\epsilon_{n}}\left[\delta_{\sum_{j\ne k}\epsilon_{j},0}\right]\prod_{\ell\ne k}\frac{1}{2}\erfc{\epsilon_\ell \frac{x-y_\ell}{\sqrt{4 t}}},
	\label{eq:Ak}
\end{eqnarray}
where $\epsilon_{\ell}$ are binary variables taking values $\pm 1$.

\subsection{Comparison with a random field Ising model}

The appearance of binary variables $\epsilon_j$ in Eq.~\eqref{eq:Ak} suggests to seek a connection to an Ising model with non-interacting spins $\epsilon_j$. Such a connection indeed exists. To see it we use an identity
\begin{equation}
	\erfc{\epsilon_j \frac{x-y_j}{\sqrt{4 t}}}=\dfrac{\exp \left(-h_j \epsilon_j\right)}{\cosh( h_j)},
\end{equation}
where $h_j$ is the magnetic field acting on the spin $\epsilon_j$, defined as
\begin{equation}
	h_{j}=\arctanh\left[ \erf{\frac{x-y_{j}}{\sqrt{4 t}}} \right].
	\label{eq:hj}
\end{equation}
This leads to an expression for the amplitude
\begin{eqnarray}
	A_{k}\left[ x,\Y_{k}
	\right]
	=\dfrac{Z\left[{\bf h}_k\right]}{\prod_{j\ne k}2\cosh(h_{j})},
\end{eqnarray}
where $Z\left[{\bf h}_k\right]$ is the partition function of an Ising chain of $2n$ spins in the ensemble with total magnetization zero, with ${\bf h}_k=(h_{-n},\ldots,h_{k-1}, h_{k+1},\ldots,h_n)$. The partition function is defined as
\begin{equation}
\label{PF}
	Z\left[{\bf h}_k\right]=\sum_{\epsilon_{-n}}\cdots \sum_{\epsilon_{k-1}}\sum_{\epsilon_{k+1}} \cdots \sum_{\epsilon_{n}}\left[\delta_{\sum_{j\ne k}\epsilon_{j},0}\right]\prod_{\ell\ne k}e^{h_{\ell}\epsilon_{\ell}}.
\end{equation}
The spins do not interact with each other. In terms of the partition function \eqref{PF} the probability of tagged particle position can be written as
\begin{equation}
	\text{Prob}_{t}(x\vert \Y)=\sum_{k=-n}^{n}\frac{g_t(x\vert
	y_{k})}{\prod_{j\ne k}2\cosh(h_{j})} Z\left[{\bf h}_k\right].
	\label{eq:prob rim}
\end{equation}

In this formulation in terms of the Ising model, the quenched case where the initial particle positions $y_j$ are fixed corresponds to a quenched magnetic field. The annealed case where the $y_j$ are fluctuating corresponds to the fluctuating magnetic field. It is well known that the properties of a random field Ising model is different in the quenched and in the annealed ensemble. Then, the difference in the statistics of the tagged particle position between the quenched and the annealed initial condition can be related to the non-equivalence of the random field Ising model in the two ensembles.

\subsection{Normal diffusion}

There is a finite number of particles, $2n+1$ in our setting, diffusing on an infinite line. Therefore the tagged particle has a normal diffusion in the large time limit. This large time is set by $t\gg L^2$, since initially particles are distributed within an interval $[-L,L]$. 

In this limit, the leading behavior of the probability given by \eqref{eq:prob rim} comes from replacing $y_j=0$ for all $j$. This corresponds to the case where the magnetic field $h_{j}\equiv h=\arctanh[\erf{x/\sqrt{4t}}]$ is uniform.
The corresponding partition function is $Z[{\bf h}_k]=\frac{(2n)!}{n! n!}$. Note that it is independent of $h$ because the total magnetization vanishes.

Substituting $y_j=0$ for all $j$ into \eqref{eq:prob rim} we obtain
\begin{equation}
	\text{Prob}_{t}(x\vert \Y)\simeq g_t(x\vert 0)\left[
	\frac{1}{4}\erfc{\frac{x}{\sqrt{4 t}}}\erfc{-\frac{x}{\sqrt{4 t}}}\right]^{n}\frac{(2n+1)!}{n!\,n!}~.
\end{equation}
The result does not depend on the initial positions $\Y$ of the particles, that is whether the initial state is annealed or quenched. For large $n$ the leading dependence on $n$ becomes
\begin{equation}
	\text{Prob}_{t}\left(x=\xi\sqrt{4t}\right)\simeq \frac{\sqrt{n}}{\pi \sqrt{t}}~e^{n \psi(\xi)},
\end{equation}
with $\psi(\xi)=\ln[\erfc{\xi}\erfc{-\xi}]$. At large times, the distribution is expected to be Gaussian. This can be confirmed by expanding $\psi(\xi)$ for small $\xi$ to give
\begin{equation}
	\text{Prob}_{t}(x)\simeq \frac{1}{\sqrt{4\pi
	\mathcal{D}t}}~e^{\displaystyle -\frac{x^2}{4\mathcal{D}t}},
\end{equation}
with the self-diffusion constant $\mathcal{D}=\frac{ \pi}{4 n}$. The self-diffusion constant decreases as $n$ increases indicating the sub-diffusive behavior observed in the macroscopic calculation.

\subsection{Sub-diffusion \label{sec:ldf single-file}}

This limit is defined by the number of particles $n\rightarrow
\infty$ and $L\rightarrow \infty$ keeping the density $n/L=\rho$ constant. The central (tagged) particle is caged by infinitely many particles. From the hydrodynamic result in earlier sections it is expected that in this limit the probability of the tagged particle position has the large deviation form $\text{Prob}_{t}(x)\asymp \exp\left[ -\sqrt{4 t}~\phi(x/\sqrt{4 t}) \right]$. We now derive $\phi(\xi)$ for quenched and annealed settings, and compare with the results from the hydrodynamic approach.

\subsubsection*{Quenched Case}

As a quenched initial state we choose the equidistant one: All particles at $t=0$ are placed deterministically with separation $\rho^{-1}$ between adjacent particles. Then the initial position of the $\ell^\text{th}$ particle is $y_{\ell}=\rho^{-1}\ell$. We write the probability given by Eq.~\eqref{eq:prob exact 0} as 
\begin{equation}
	\text{Prob}_t(x)=P_0(x)+P_{-}(x)+P_{+}(x),
	\label{eq:prob exact}
\end{equation}
with
\begin{eqnarray}
	P_0(x) &&= g_t(x\vert 0)A_{0}\left[ x, \mathbf{Y}_{0}\right], \\
	P_{-}(x) &&= \sum_{k=-1}^{-n} g_t(x\vert y_{k})A_{k}\left[ x, \mathbf{Y}_{k}\right], \\
	P_{+}(x) &&=  \sum_{k=1}^{n} g_t(x\vert y_{k})A_{k}\left[ x, \mathbf{Y}_{k}\right].
\end{eqnarray}
All the three terms $P_{0}$, $P_{\pm}$ have the same asymptotic large deviation form at large $t$ limit. They only differ in the sub-leading terms in $t$. The advantage of using the representation \eqref{eq:prob exact} is that it is much simpler to analyze the term $P_{0}$ and to extract the large deviation function. We derive the large deviation function using a saddle point analysis of $P_{0}$.

We use the expression for $A_0$ from \eqref{eq:Ak} which leads to
\begin{equation}
P_{0}(x)=g_t(x \vert 0)\sum_{\epsilon_{-n}}\cdots \sum_{\epsilon_{-1}}\sum_{\epsilon_{1}} \cdots \sum_{\epsilon_{n}}\left[\delta_{\sum_{j\ne 0}\epsilon_{j},0}\right]\prod_{\ell\ne 0}\frac{1}{2}\erfc{\epsilon_\ell \frac{x-\rho^{-1}\ell}{\sqrt{4 t}}}.
\end{equation}
We have used the quenched initial particle position $y_{\ell}=\rho^{-1}\ell$. The Kronecker delta function can be replaced by an integral,
\begin{equation*}
P_{0}(x)=g_t(x \vert 0)\int dB~\sum_{\epsilon_{-n}}\cdots \sum_{\epsilon_{-1}}\sum_{\epsilon_{1}} \cdots \sum_{\epsilon_{n}}\prod_{\ell\ne 0}\left[e^{\epsilon_{\ell} B/2}\right]\frac{1}{2}
\erfc{\epsilon_\ell \frac{x-\rho^{-1}\ell}{\sqrt{4 t}}}\,,
\end{equation*}
where $B$ is the integration variable and the factor $1/2$ is for later convenience. In this form, the binary variables $\epsilon_{\ell}$ are decoupled and their sums can be performed rather easily, leading to
\begin{eqnarray}
P_{0}(x)&&=g_t(x \vert 0)\int dB~\exp\left[\sum_{\ell=1}^n\log\left\{1+\left(e^{B}-1\right)\frac{1}{2}\erfc{ \frac{x+\rho^{-1}\ell}{\sqrt{4 t}}}\right\}\right.\nonumber\\
&&\qquad \qquad \qquad \qquad \left. +\sum_{\ell=1}^n\log\left\{1+\left(e^{-B}-1\right)\frac{1}{2}\erfc{ \frac{\rho^{-1}\ell-x}{\sqrt{4 t}}}\right\}  \right] \nonumber.
\end{eqnarray}
Taking the limit $n\rightarrow \infty$ with uniform density of particles $\rho$ and replacing the summation over $\ell$ by an integral, yields, 
\begin{eqnarray}
P_{0}(x)&&=g_t(x \vert 0)\int dB~\exp\left[\rho \int_{0}^{\infty}dz\log\left\{1+\left(e^{B}-1\right)\frac{1}{2}\erfc{ \frac{x+z}{\sqrt{4 t}}}\right\}\right.\nonumber\\
&&\qquad \qquad \qquad \qquad \left.+ \rho \int_{0}^{\infty}dz\log\left\{1+\left(e^{-B}-1\right)\frac{1}{2}\erfc{ \frac{z-x}{\sqrt{4 t}}}\right\}  \right] \nonumber.
\end{eqnarray}

To compute the large deviation function we write $\xi=x/\sqrt{4 t}$ and recast the above formula into 
\begin{eqnarray}
P_0(\xi\sqrt{4t})&&=\dfrac{\exp(-\xi^2)}{\sqrt{4\pi t}}\int dB~\exp\left[\rho \sqrt{4 t} \int_{\xi}^{\infty}dz\log\left\{1+\left(e^{B}-1\right)\frac{1}{2}\erfc{ z}\right\}\right.\nonumber\\
&&\qquad \qquad \qquad \qquad \qquad \left.+ \rho  \sqrt{4 t} \int_{-\xi}^{\infty}dz\log\left\{1+\left(e^{-B}-1\right)\frac{1}{2}\erfc{z}\right\}  \right] \nonumber.
\end{eqnarray}
At large $t$ the integral over $B$ is dominated by the saddle point leading to a parametric solution of the large deviation function 
\begin{eqnarray}
\phi_{ \mathcal{\scriptstyle Q}}(\xi) = - \lim_{t\rightarrow\infty}\frac{\log P_0\left(\xi\sqrt{4t}\right)}{\sqrt{4 t}}&&=
- \rho  \int_{\xi}^{\infty}dz\log\left\{1+\left(e^{B}-1\right)\frac{1}{2}\erfc{ z}\right\}\nonumber\\
&& -  \rho   \int_{-\xi}^{\infty}dz\log\left\{1+\left(e^{-B}-1\right)\frac{1}{2}\erfc{z}\right\} 
\end{eqnarray}
with $B$ determined from the optimization requirement $	\frac{d\phi_{\mathcal{\scriptstyle Q}}(\xi)}{dB}=0$.
These results are identical to \eqref{eq:y Q final}--\eqref{eq:ldf Q alternate form} obtained using the MFT.

\subsubsection*{Annealed Case}

In this case, the initial positions of the particles are uniformly distributed in the interval $[-L, L]$. Averaging over the initial positions the probability of the tagged particle position can be expressed as
\begin{align}
	\text{Prob}_t(x)&=\left[\frac{n!}{L^{n}}\int_{-L}^{0}dy_{-1}\int_{-L}^{y_{-1}}dy_{-2}\cdots \int_{-L}^{y_{-n+1}}dy_{-n}\right]\nonumber\\
	&\qquad\qquad \qquad \left[\frac{n!}{L^{n}}\int_{0}^{L}dy_{1}\int_{y_1}^{L}dy_{2}\cdots \int_{y_{n-1}}^{L}dy_{n}\right] \text{Prob}_t(x\vert \Y),
	\label{eq:Annealed Micro step 0}
\end{align}
where $\text{Prob}_t(x\vert \Y)$ is given in \eqref{eq:prob exact 0}. As in the quenched case, we write $\text{Prob}_t(x)$ as a sum of three terms. All terms have the same large deviation form. Hence it is sufficient to analyze 
\begin{align*}
	\text{Prob}_t(x)\sim P_0(x)= g_t(x\vert 0)
	&\Bigg[\frac{n!}{L^{n}}\int_{-L}^{0} dy_{-1}\int_{-L}^{y_{-1}}dy_{-2}\cdots \int_{-L}^{y_{-n+1}}dy_{-n}\Bigg]\\
	& \times \left[\frac{n!}{L^{n}}\int_{0}^{L}dy_{1}\int_{y_1}^{L}dy_{2}\cdots \int_{y_{n-1}}^{L}dy_{n}\right]\\
	& \times \sum_{\epsilon_{-n}}\cdots \sum_{\epsilon_{-1}}\sum_{\epsilon_{1}} \cdots \sum_{\epsilon_{n}}\left[\delta_{\sum_{j\ne 0}\epsilon_{j},0}\right]\prod_{\ell\ne 0}\frac{1}{2}\erfc{\epsilon_\ell \frac{x-y_{\ell}}{\sqrt{4 t}}}.
\end{align*}

The symmetry of the integrand allows us to re-write it as
\begin{align*}
	P_0(x) = g(x\vert 0)&
	\sum_{\epsilon_{-n}}\cdots \sum_{\epsilon_{-1}}\sum_{\epsilon_{1}} \cdots \sum_{\epsilon_{n}}\left[\delta_{\sum_{j\ne 0}\epsilon_{j},0}\right]\\
	&\left[\prod_{\ell=-1}^{-n}\frac{1}{L}\int_{-L}^{0}dy_{\ell}\frac{1}{2}\erfc{\epsilon_{\ell} \frac{x-y_{\ell}}{\sqrt{4 t}}} \right]
	\left[ \prod_{\ell=1}^{n} \frac{1}{L}\int_{0}^{L}dy_{\ell}\,\frac{1}{2}\erfc{\epsilon_\ell \frac{x-y_{\ell}}{\sqrt{4 t}}}\right].
\end{align*}

The large deviation function can be derived from the above expression using a saddle point approximation. Replacing the delta function by an integral representation we get
\begin{align*}
	P_0(x) =  g_t(x\vert 0)\int dB & \left[\prod_{\ell=-1}^{-n} 
	\sum_{\epsilon_{\ell}}e^{\epsilon_{\ell} B/2}\frac{1}{L}\int_{-L}^{0}dy_{\ell}\frac{1}{2}\erfc{\epsilon_{\ell} \frac{x-y_{\ell}}{\sqrt{4 t}}} \right]\\
	&\left[ \prod_{\ell=1}^{n} \sum_{\epsilon_{\ell}}e^{\epsilon_{\ell} B/2} \frac{1}{L}\int_{0}^{L}dy_{\ell}\frac{1}{2}\erfc{\epsilon_\ell \frac{x-y_{\ell}}{\sqrt{4 t}}}\right].
\end{align*}
Completing the summation over the binary variables $\epsilon_{\ell}$ we get
\begin{align}
	P_0(x) = g_t(x\vert 0)\int dB & \left[
	1+\left(e^{B}-1\right)\frac{1}{L}\int_{-L}^{0}dy\frac{1}{2}\erfc{ \frac{x-y}{\sqrt{4 t}}} \right]^{n}\nonumber\\
	&\left[  1+\left(e^{-B}-1\right) \frac{1}{L}\int_{0}^{L}dy\frac{1}{2}\erfc{ \frac{y-x}{\sqrt{4 t}}}\right]^{n}\nonumber,
\end{align}
where we have replaced $y_\ell$ by $y$, as the former reduces to a dummy variable after the summation.

In the limit of $n\rightarrow \infty$ and $L\rightarrow \infty$ keeping $n/L=\rho$ fixed, the above expression results in
\begin{equation*}
	P_0(x)= g_t(x\vert 0)\int dB \exp\left[
	\rho \left(e^{B}-1\right)\int_{-\infty}^{0}dy\frac{1}{2}\erfc{ \frac{x-y}{\sqrt{4 t}}}  + \rho\left(e^{-B}-1\right) \int_{0}^{\infty}dy\frac{1}{2}\erfc{ \frac{y-x}{\sqrt{4 t}}}\right].
\end{equation*}
In terms of rescaled coordinates $\xi=x/\sqrt{4 t}$ and $z=y/\sqrt{4 t}$ the above can be written as
\begin{equation*}
	P_0(\xi\sqrt{4 t})= \dfrac{e^{-\xi^2}}{\sqrt{4\pi t}}\int dB \exp\left[
	\rho \sqrt{t}\left\{\left(e^{B}-1\right)\int_{-\infty}^{0}dz~\erfc{ \xi-z} +\left(e^{-B}-1\right) \int_{0}^{\infty}dz~\erfc{ z-\xi}\right\}\right].
\end{equation*}
At large $t$, the integral is dominated by the saddle point. This leads to the large deviation function
\begin{equation}
\phi_{ \mathcal{\scriptstyle A}}(\xi) \simeq - \lim_{t\to \infty}\frac{\log P_0\left(\xi\sqrt{4t}\right)}{\sqrt{4t}}
= - \rho  \frac{\left(e^{B}-1\right)}{2}\int_{\xi}^{\infty}dz~\erfc{ z}
  - \rho \frac{\left(e^{-B}-1\right)}{2}\int_{-\xi}^{\infty}dz~\erfc{z},
 \label{eq:phi Ann micro}
\end{equation}
with $B$ determined from the saddle point condition
\begin{equation}
	\frac{d\phi_{\mathcal{\scriptstyle A}}(\xi)}{dB}=0.
	\label{eq:B Ann micro}
\end{equation}

To show the equivalence with the MFT result \eqref{eq:ldf final} we use \eqref{eq:B Ann micro} to deduce 
\begin{equation}
e^{2B}=\frac{ {\displaystyle \int_{-\xi}^{\infty}} dz~\erfc{z}}{{\displaystyle\int_{\xi}^{\infty}}dz~\erfc{z}}.
\end{equation}
Combining this with \eqref{eq:phi Ann micro} one recovers the large deviation function \eqref{eq:ldf final}.

\section{Summary}

We studied the full statistics of the displacement of the tagged particle in single-file diffusion. Our analysis mostly relies on the macroscopic fluctuation theory (MFT). We found the full solution for the simplest single-file system composed of impenetrable Brownian particles (we computed the optimal paths, the large deviation function, etc.). This single-file system is also amenable to an exact microscopic analysis. In the annealed case, the large deviation function was originally computed in \cite{Rodenbeck1998} and we presented another (shorter) derivation; the quenched case has not been studied before. The predictions based on the exact analyses and those which were derived using the MFT  match for the single-file system composed of Brownian particles. 

Single-file systems with arbitrary transport coefficients cannot be solved exactly. Yet such general systems are tractable perturbatively. Specifically, we performed an expansion in the powers of the fugacity and derived an exact formula for the variance of position of the tagged particle valid in the general case of arbitrary $D(\rho)$ and $\sigma(\rho)$. The forth cumulant can also be computed for a fairly general class of single-file systems, e.g., for systems with constant diffusion coefficient and quadratic (in density) mobility. The well-known example of such system is the SEP, and we computed the forth cumulant of the tagged particle position for the SEP.

Our study can be extended in many directions. One should be able  to use the MFT to calculate statistical properties of the tagged particle trajectory (the two-time correlation functions \cite{KMS2015}, the first passage times etc.). It would also be interesting to investigate the influence of a bias in the simple exclusion process. 
If one considers an asymmetrically hopping tagged particle in a bath of SEP,  all the  cumulants scale as $\sqrt{T}$ \cite{Burlatsky1992,Landim1998}. The full statistics has been recently determined in the high density limit  \cite{Illien2013} when the leading asymptotic can be extracted by treating the vacancies as independent random walkers; for the finite density even the variance remains unknown. If all particles are biased, even the scale of fluctuations depend on the setting: In the annealed case, the position fluctuation of the tagged particle grows as $\sqrt{T}$ \cite{Kipnis1986,Liggett2004}; in the quenched case, the exponent changes from $1/2$ to $1/3$ \cite{Demasi1985,ShamikSatya,Beijeren1991,Presutti,Ferrari91,Alexander93}. It is not known whether a hydrodynamic approach based on the MFT  can capture these behaviors and if the Tracy-Widom distribution,  which is ubiquitous in these problems (see e.g. \cite{Imamura2007}), can be retrieved as a solution of some optimal path  equations.

From a more general  point of view, it would be interesting to give a physical interpretation of
 the conjugate field that appears in the   MFT  optimal path equations  and 
 to classify the one-dimensional gases that could lead to  classically integrable
 partial differential  equations,  that can be solved using the inverse scattering method \cite{Babelon}.

\begin{acknowledgements}
We benefitted from discussions with A. Dhar,  S. Majumdar, S. Mallick, B. Meerson,  S. Sabhapandit  and R. Voituriez. We are also grateful to the referees for very useful comments and suggestions.This research was partly supported by grant No.\ 2012145 from the BSF. We thank the Galileo Galilei Institute for Theoretical Physics for excellent working conditions and the INFN for partial support. 
\end{acknowledgements}

\appendix

\section{Functional derivative of $Y[q]$ \label{sec:functional derivative}}

The tagged particle position is a functional $Y[q]$ of the initial $q(x,0)$ and the final $q(x,T)$ density profile, defined by the relation
\begin{equation}
	\int_{0}^{Y[q]}dx ~q(x,T)=\int_{0}^{\infty}dx~\bigg[ q(x,T)-q(x,0) \bigg].
	\label{eq:without variation}
\end{equation}
which comes from  equation
 (\ref{eq:X_T def}). This way of writing makes both the integrals convergent.

The functional derivatives appearing in the boundary conditions \eqref{eq:pxT}--\eqref{eq:px0 annealed} can be easily computed from the definition above. To proceed, consider a small variation in the final density $q(x,T)\rightarrow q(x,T)+\delta q(x,T)$ leading
to a change $Y\rightarrow Y+\delta Y$. Corresponding to this variation the above equation becomes
\begin{equation*}
	\int_{0}^{Y+\delta Y}dx \bigg[q(x,T)+\delta q(x,T)\bigg]=\int_{0}^{\infty}dx~\bigg[ q(x,T)-q(x,0)
	\bigg]+\int_{0}^{\infty}dx ~\delta q(x,T).
\end{equation*}
Using the formula \eqref{eq:without variation} this reduces to,
\begin{equation}
	\int_{Y}^{Y+\delta Y}dx ~q(x,T)=\int_{Y+\delta Y}^{\infty}
	dx~\delta q(x,T) \, .
\end{equation}
Assuming that the change $\delta Y$ is small for a small variation $\delta q(x,T)$ and keeping only the linear terms, the above leads to
\begin{equation}
	\delta Y=\int_{-\infty}^{\infty}dx
	\left[\frac{\Theta(x-Y)}{q(Y,T)}\right]\delta q(x,T) \, .
\end{equation}
Hence, we have 
\begin{equation}
	\frac{\delta Y}{\delta q(x,T)}=\frac{\Theta(x-Y)}{q(Y,T)}.
\end{equation}
The functional derivative with respect to the initial density $q(x,0)$ is similarly derived:
\begin{equation}
	\frac{\delta Y}{\delta q(x,0)}=-\frac{\Theta(x)}{q(Y,T)}.
\end{equation}

\section{Inequality between cumulants in the annealed and quenched cases \label{app:inequality}}

Because of the fluctuations in the initial state, the tagged particle is expected to have larger displacements in the annealed case than in the quenched case. More generally for any single file system
\begin{equation}
 \mu_{\mathcal{A}}(\lambda) \ge \mu_{\mathcal{Q}}(\lambda)
\label{eq:inequality}
\end{equation}
which implies $\left \langle X_{T}^{2} \right\rangle_{\mathcal{A}} \ge \left \langle X_{T}^{2} \right\rangle_{\mathcal{Q}}$.

To derive \eqref{eq:inequality} we recall that the two initial settings differ by how the average over initial state is taken:
\begin{eqnarray}
\mu_{\mathcal{A}}(\lambda)& =&\log \left \langle e^{\lambda X_T} \right\rangle_{\textrm{evolution+initial}},\\
\mu_{\mathcal{Q}}(\lambda)&=&\left \langle\log \left\langle e^{\lambda X_T} \right\rangle_{\textrm{evolution}}\right\rangle_{\textrm{initial}}.
\end{eqnarray}
Here the subscript  {\it  evolution}  denotes average over stochastic evolution of the system and  the subscript {\it initial}  denotes average over initial state. Equation \eqref{eq:inequality} then follows  from the Jensen inequality \cite{Gradshteyn2007} because $\log$ is a concave function.   The  inequality  \eqref{eq:inequality}  
implies the opposite relation for the large deviation functions and therefore explains why the large deviation function in the quenched case exceeds the large deviation function in the annealed case (Fig.~\ref{fig:ldf}). 

\section{ Mobility $\sigma(\rho)$ \label{sec:sigma derivation}}

We first derive $\sigma(\rho)$ for the Brownian particles with hard core repulsion. Let there are $n$ number of particles at equilibrium within an interval of length $L$. All particle positions within the interval are equally probable. This leads to the canonical free energy density
\begin{equation*}
f\left(\dfrac{n}{L} \right)=-L^{-1}\log\bigg( \frac{L^n}{n!} \bigg).
\end{equation*}
In the limit $n\rightarrow\infty$ and $L\rightarrow\infty$ with finite $n/L=\rho$, the above formula becomes 	
$f(\rho)=\rho \ln \rho -\rho$. Using it together with the fluctuation-dissipation relation \eqref{eq:fluc diss} 
one gets $\sigma(\rho)=\frac{2D(\rho)}{f^{\prime\prime}(\rho)}=2\rho$.

Consider now the SEP with $n$ particles on the ring of $L$ sites. In the equilibrium, all configurations are equally probable which leads to the free energy density
\begin{equation*}
f\left(\dfrac{n}{L} \right)=-L^{-1}\log{L \choose n}.
\end{equation*}
Using Stirling's  approximation  we get $f(\rho)=\rho \ln \rho +( 1-\rho)\ln( 1-\rho)$. Using the fluctuation-dissipation relation we recover the well-known formula $\sigma(\rho)=2\rho( 1-\rho)$ for the mobility in the case of SEP.

\section{Derivation of the integral \eqref{eq:integral one} \label{sec:current}}

The result in \eqref{eq:integral one} can be proved by comparing with the analysis of the time integrated current in a symmetric exclusion process on an infinite line \cite{Gerschenfeld2009Bethe,Gerschenfeld2009}. Consider an annealed initial state at uniform density $\rho$. Let $Q_T$ be the time integrated current in a time window $[0,T]$ through the site at origin. The current is related to the hydrodynamic density profile by 
\begin{equation*}
Q_T={\displaystyle\int_{0}^{\infty}}dx\Bigg[\rho(x,T)-\rho(x,0)\Bigg].
\end{equation*}

On an average the current is zero because of the uniform initial density. However, all the even cumulants are non-zero. The analysis for the cumulant generating function of $Q_T$ can be easily formulated in terms of the macroscopic fluctuation theory \cite{Gerschenfeld2009}. To analyze the variational formulation we take a series expansion method, similar to the one presented in \sref{sec:fourth}. This way the fourth cumulant of current for the symmetric exclusion process can be related to that for non-interacting particles. The relation is similar to the one in \eqref{eq:mapping one}:
\begin{equation}
 \left  \langle Q_T^4 \right \rangle_c=(1-\alpha\rho)(1-2\alpha\rho)^2 
\left \langle \widehat{Q}_T^4 \right \rangle_c+   24 \, \alpha  \rho^2(1-\alpha\rho)^2
\Bigg[ - \mathcal{I}+\int_{0}^{\infty}dx\Bigg(\dfrac{{h}(x,T)}{\rho^2}-\dfrac{{h}(x,0)}{\rho^2}\Bigg)\Bigg],
\label{eq:fourth current}
\end{equation}
where $\mathcal{I}$ and $\widehat{h}$ are defined in section \ref{sec:fourth}. We followed the same convention of denoting the non-interacting case by hat variables.

Using the Bethe ansatz Derrida and Gerschenfeld derived an exact expression for all the cumulants of the current for the SEP in the annealed setting \cite{Gerschenfeld2009Bethe}. The fourth cumulant in the annealed case can be extracted from their work:
\begin{equation}
\left  \langle Q_T^4\right  \rangle_c=\rho(1-\rho)\Bigg[1-3\sqrt{2}\rho(1-\rho)\Bigg]\dfrac{2}{\sqrt{\pi}}.
\end{equation}
On the other hand the cumulant for non-interacting particles corresponding to $\alpha=0$ can be derived by
 a direct  counting  argument \cite{Gerschenfeld2009} leading to 
\begin{equation}
\left \langle {\widehat{Q}}_T^4 \right\rangle_c=\rho~\dfrac{2}{\sqrt{\pi}}.
\end{equation}
Making use of these results we establish \eqref{eq:integral one}.

\section{Function $G$ \label{app:identity 1 G}}

Here we prove an identity for the function $G$ defined in equation \eqref{eq:G def}. Let $\gamma$ be a
permutation of the set $\Omega=\left\{y_{-n},\cdots,y_{-1},y_{1},\cdots,y_{n}  \right\}$ of $2n$ elements. Note that the zeroth element has been excluded for convenience. The function $G$ has the property that
\begin{align}
\sum_{\gamma}G\left(
	-x,-y_{\gamma(-n)},\cdots,-y_{\gamma(-2)},-y_{\gamma(-1)}
	\right)
	& G\left(x,y_{\gamma(1)},y_{\gamma(2)},\cdots,y_{\gamma(n)}\right)\nonumber\\
	&=\sum_{\epsilon_{-n}}\cdots \sum_{\epsilon_{-1}}\sum_{\epsilon_{1}} \cdots \sum_{\epsilon_{n}}\left[\delta_{\sum_{j\ne 0}\epsilon_{j},0}\right]  \prod_{\ell\ne 0}\frac{1}{2}\erfc{\epsilon_\ell \frac{x-y_\ell}{\sqrt{4 t}}}.
	\label{eq:id 0}
\end{align}
The identity holds for any set of $2n$ elements, with $n$ being a positive integer, and the binary variables $\epsilon_j=\pm 1$ for any $j$.

First, we want to prove
\begin{equation}
	\sum_{\nu}G(x,y_{\nu(1)},\cdots,y_{\nu(n)})=\prod_{j=1}^{n}\frac{1}{2}\erfc{\frac{x-y_{j}}{\sqrt{4 t}}},
	\label{eq:id 2}
\end{equation}
where $\nu$ is a permutation of the set $\{y_1,\cdots,y_n \}$. Using the definition of $G$ in \eqref{eq:G def} we get
\begin{eqnarray*}	
	\sum_{\nu}G(x,y_{\nu(1)},\cdots,y_{\nu(n)})=\sum_{\nu}\int_{x}^{\infty}dz_{1}	\int_{z_{1}}^{\infty}dz_{2}\cdots\int_{z_{n-1}}^{\infty}dz_{n}
	\prod_{j=1}^{n} g_t\left(z_{j}\vert y_{\nu(j)} \right).
\end{eqnarray*}
As $z_{j}$'s are dummy variables, the expression can be re-written as
\begin{equation*}	
	\sum_{\nu}G(x,y_{\nu(1)}
	\cdots, y_{\nu(n)})
=\left[\sum_{\nu}\int_{x}^{\infty}dz_{\nu(1)}
\int_{z_{\nu(1)}}^{\infty}dz_{\nu(2)}\cdots
\int_{z_{\nu(n-1)}}^{\infty}dz_{\nu(n)} \right]
\prod_{j=1}^{n}g_t\left(z_{j}\vert y_j \right).
\end{equation*}
The range of integration denoted inside the square bracket is equal to the integration over the entire volume of all $z_{j}\ge x$, leading to
\begin{equation}	
\label{ident:3}
	\sum_{\nu}G(x,y_{\nu(1)},\cdots,y_{\nu(n)}) =\prod_{j=1}^{n}\left[\int_{x}^{\infty}dz_{j}~g_t\left(z_{j}\vert
	y_{j} \right)\right]
 =\prod_{j=1}^{n}\frac{1}{2}\erfc{\frac{x-y_j}{\sqrt{4t}}},
\end{equation}
proving \eqref{eq:id 2}.

In the next step we prove the identity \eqref{eq:id 0} using \eqref{ident:3}. In \eqref{eq:id 0} the summation is over all permutations of the set $\Omega$. Any ordered set which is generated by a permutation of $\Omega$ can also be generated by suitably constructing two subsets of $n$ elements each, and subsequently taking permutations within the elements of the subsets. To elaborate, let us divide the ordered set $\Omega\equiv\{y_{-n},\cdots,y_{-1},y_{1},\cdots,y_{n} \}$ into two ordered subsets $\omega_{L}\equiv\{y_{-n},\cdots,y_{-1} \}$ and $\omega_{R}\equiv\{y_{1},\cdots,y_{n} \}$. The number of ordered sets generated by permutations of $\Omega$ is $(2n)!$, all of which can also be generated by combination of the following steps.
\begin{enumerate}
\item 
The simplest are permutations strictly within $\omega_{L}$ and $\omega_{R}$. There are $n!n!$ such permutations.
\item
One can construct another pair of ordered subsets $\omega_{L}^{\prime}$ and $\omega_{R}^{\prime}$ by exchanging any two elements between $\omega_{L}$ and $\omega_{R}$. There are $n^2$ pair of subsets that can be generated this way. Overall this produces $n!n! {n \choose 1}^2$ permutations.
\item
We can further exchange two elements from $\omega_{L}$ with two elements from $\omega_{R}$, and then consider permutations within those subsets. This generates  $n!n! {n \choose 2}^2$ permutations.
\item
Continuing this process we can generate all permutations $\Omega$.
\end{enumerate}
As a useful check, one can compute all the permutations generated by the above procedure and find that it is equal to the total number of permutations:
\begin{equation*}
n!n!\left[1+{n \choose 1}^2+{n \choose 2}^2+\cdots+{n \choose n}^2\right]=(2n)!
\end{equation*}

The above decomposition of permutations simplifies the summation on the left hand side in \eqref{eq:id 0}. For example, consider the ordered sets generated in step 1 of the decomposition where the ordered subsets are $\nu_1\omega_{L}$ and $\nu_2\omega_{R}$, with $\nu_1$ and $\nu_2$ being permutation operators. Summing over all permutations we have
\begin{equation*}
\sum_{\nu_1}G(-x,-y_{\nu_1(-n)},\cdots,-y_{\nu_1(-1)})\sum_{\nu_2}G(x,y_{\nu_2(1)},\cdots,y_{\nu_2(n)})=\prod_{j=1}^{n}\frac{1}{2}\erfc{-\frac{x-y_{-j}}{\sqrt{4t}}}\frac{1}{2}\erfc{\frac{x-y_j}{\sqrt{4t}}},
\end{equation*}
where we used the relation \eqref{eq:id 2}. We now introduce binary variables $\{ \epsilon_{-n},\cdots,\epsilon_{-1},\epsilon_{1},\cdots,\epsilon_{n}\}$, each of which take values $\pm 1$. Then the above formula can be re-written as
\begin{equation*}
\sum_{\nu_1}G(-x,-y_{\nu_1(-n)},\cdots,-y_{\nu_1(-1)})\sum_{\nu_2}G(x,y_{\nu_2(1)},\cdots,y_{\nu_2(n)})=\prod_{j=-n, j\ne 0}^{n}\frac{1}{2}\erfc{\epsilon_{j}\frac{x-y_j}{\sqrt{4t}}},
\end{equation*}
with a configuration of the binary variables: $\epsilon_j=-1$ for $j<0$ and  $\epsilon_j=1$ for $j>0$.

The summation over configurations generated in the step 2 is performed in a similar way. In this, the subsets are generated by exchange of one element each from $\omega_L$ and $\omega_R$. For example, consider the $(-n)$th element from $\omega_L$ and $1$st element from $\omega_R$ have been exchanged. The summation over configurations generated by permutation within these subsets yields a simple expression
\begin{equation*}
\sum_{\nu_1}G(-x,-y_{\nu_1(1)},-y_{\nu_1(-n+1)}\cdots,-y_{\nu_1(-1)}) \sum_{\nu_2}G(x,y_{\nu_2(-n)},y_{\nu_2(2)},\cdots,y_{\nu_2(n)})=\prod_{j=-n, j\ne 0}^{n}\frac{1}{2}\erfc{\epsilon_{j}\frac{x-y_j}{\sqrt{4Dt}}},
\end{equation*}
with $\epsilon_{-n}=1$, $\epsilon_{1}=-1$ and other binary variables are the same as before. From these examples the pattern emerges, namely the above decomposition of permutations leads to the identity \eqref{eq:id 0}.


\bibliographystyle{iopart-num}
\bibliography{reference}

\end{document}